\documentclass[aps,prb,twocolumn,superscriptaddress,floatfix,amsmath,amssymb,longbibliography]{revtex4-2}

\usepackage{physics}
\usepackage{graphicx}
\usepackage{bm}
\usepackage{amssymb}
\usepackage{amsmath}
\usepackage{hyperref}
\usepackage{dsfont}
\usepackage{mathtools}
\usepackage{tikz-cd}

\newcommand{\YoungSix}{%
\begin{tikzpicture}[scale=0.4, baseline=-0.5ex]
  \foreach \x in {0,...,5} {
    \draw (\x,0) rectangle ++(1,1);
  }
\end{tikzpicture}%
}

\newcommand{\YoungFiveOne}{%
\begin{tikzpicture}[scale=0.4, baseline=-0.5ex]
  \foreach \x in {0,...,4} {
    \draw (\x,0) rectangle ++(1,1);
  }
  \draw (0,-1) rectangle ++(1,1);
\end{tikzpicture}%
}

\newcommand{\YoungFourTwo}{%
\begin{tikzpicture}[scale=0.4, baseline=-0.5ex]
  \foreach \x in {0,...,3} {
    \draw (\x,0) rectangle ++(1,1);
  }
  \foreach \x in {0,...,1} {
    \draw (\x,-1) rectangle ++(1,1);
  }
\end{tikzpicture}%
}

\newcommand{\YoungThreeThree}{%
\begin{tikzpicture}[scale=0.4, baseline=-0.5ex]
  \foreach \x in {0,...,2} {
    \draw (\x,0) rectangle ++(1,1);
  }
  \foreach \x in {0,...,2} {
    \draw (\x,-1) rectangle ++(1,1);
  }
\end{tikzpicture}%
}

\newtheorem{definition}{Definition}
\newtheorem{theorem}{Theorem}

\begin{document}

\title{Classical vs quantum dynamics and the onset of chaos in a macrospin system}

\author{Haowei Fan}
\affiliation{Department of Physics, City University of Hong Kong, Kowloon, Hong Kong SAR}
\affiliation{Department of Physics and Astronomy, University of Manchester, Oxford Road, Manchester, M13 9PL, United Kingdom}
\affiliation{National Graphene Institute, University of Manchester, Booth St.\ E., Manchester, M13 9PL, United Kingdom}

\author{Vladimir Fal'ko}
\affiliation{Department of Physics and Astronomy, University of Manchester, Oxford Road, Manchester, M13 9PL, United Kingdom}
\affiliation{National Graphene Institute, University of Manchester, Booth St.\ E., Manchester, M13 9PL, United Kingdom}
\email{vladimir.falko@manchester.ac.uk}

\author{Xiao Li}
\affiliation{Department of Physics, City University of Hong Kong, Kowloon, Hong Kong SAR}
\email{xiao.li@cityu.edu.hk}

\date{\today}%

\begin{abstract}
We study a periodically driven macrospin system with anisotropic long-range interactions and collective dissipation, described by a Lindblad master equation. In the thermodynamic limit ($N\to\infty$), a mean-field treatment yields classical equations of motion, whose dynamics are characterized via the maximal Lyapunov exponent (MLE). Focusing on the thermodynamic limit, we map out chaotic, quasiperiodic, and periodic phases via bifurcation diagrams, MLEs, and Fourier spectra of evolved observables, identifying classic period-doubling bifurcations and fractal boundaries in the regions of attractors. Finite-size quantum simulations in the Dicke basis reveal that while both quantum and classical systems exhibit diverse dynamical phases, finite-size effects suppress some behaviors present in the thermodynamic limit. The sign of $\lambda_{\mathrm{max}}$ serves as a key indicator of convergence between quantum and classical dynamics, which agree over timescales up to the Lyapunov time. Analysis of the density matrix shows that convergence occurs only when its nonzero elements are sharply localized. However, the nonconvergence does not imply a fundamental difference between quantum and classical dynamics: in chaotic regimes, although the evolution orbits of quantum and classical systems show significant differences, quantum evolution becomes mixed and diffusively explores the Hilbert space, signaling quantum chaos, which can be confirmed by the delocalized nature of the density matrix.

\end{abstract}
\maketitle


\section{Introduction}
Periodically driven quantum systems have emerged as a versatile platform for engineering novel phases of matter far from equilibrium. Under suitable conditions, such systems—known as Floquet systems—can exhibit phenomena unattainable in equilibrium settings, including nonequilibrium phase transitions~\cite{PhysRevLett.95.260404, PhysRevLett.102.100403,PhysRevLett.108.043003}, topologically nontrivial systems~\cite{PhysRevB.79.081406,PhysRevB.82.235114,lindner2011floquet,jotzu2014experimental,aidelsburger2015measuring}, and discrete time crystals (DTCs)~\cite{PhysRevResearch.5.L032024,PhysRevLett.117.090402,PhysRevX.7.011026,PhysRevLett.118.030401}. In experimental platforms, such as ultracold atoms, trapped ions, or superconducting qubits, the presence of periodic drive is often accompanied by coupling to the environment, where dissipation determines the system's long-time behavior~\cite{PhysRevE.55.300,PhysRevLett.117.250401,hartmann2017asymptotic,PhysRevE.91.030101,shirai2016effective,PhysRevLett.123.120602}. Dissipative Floquet systems can exhibit steady states that result from a nontrivial balance between external driving, many-body interactions, and environmental damping~\cite{PhysRevA.103.013306}. 

Recent works have revealed that such systems can support rich dynamical phenomena such as the dissipative DTC~\cite{harmon2025collectiveexcitationsdissipativetime,haga2025timeglassessymmetrybroken,jiao2024manybodynonequilibriumdynamicsselfinduced,PhysRevLett.134.223601,villa2024topologicalclassificationdrivendissipativenonlinear,PhysRevLett.134.090402,jara2025kibblezurekmechanismdissipativediscrete,sahoo2025generatingdiscretetimecrystals,arumugam2025injectionlockingrydbergdissipative}. On the one hand, the recent studies explored in detail various systems with short-range interactions or very few degrees of freedom. On the other hand, fundamental questions remain open about the systems with long-range interactions, such as cavity QED setups and trapped ions, in particular, related to their dynamics under periodic driving in the presence of dissipation. In such systems, one can separate the mean-field degree of freedom (\textbf{macrospin}) from magnon-like excitations. For multi-spin systems driven by homogeneous magnetic field, it is the macrospin that couples to the external drive, so that one may expect quantum dynamics to coincide with precession of a classical rotator. However, the quantum tunneling is capable to deviate the actual evolution of macrospin from what is expected based on the classical description. This is particularly interesting to investigate in the regime where classical dynamics undergoes transition to chaos, which will be addressed in the present study.

\begin{figure}[t]
\centering
\includegraphics[width=0.47\textwidth]{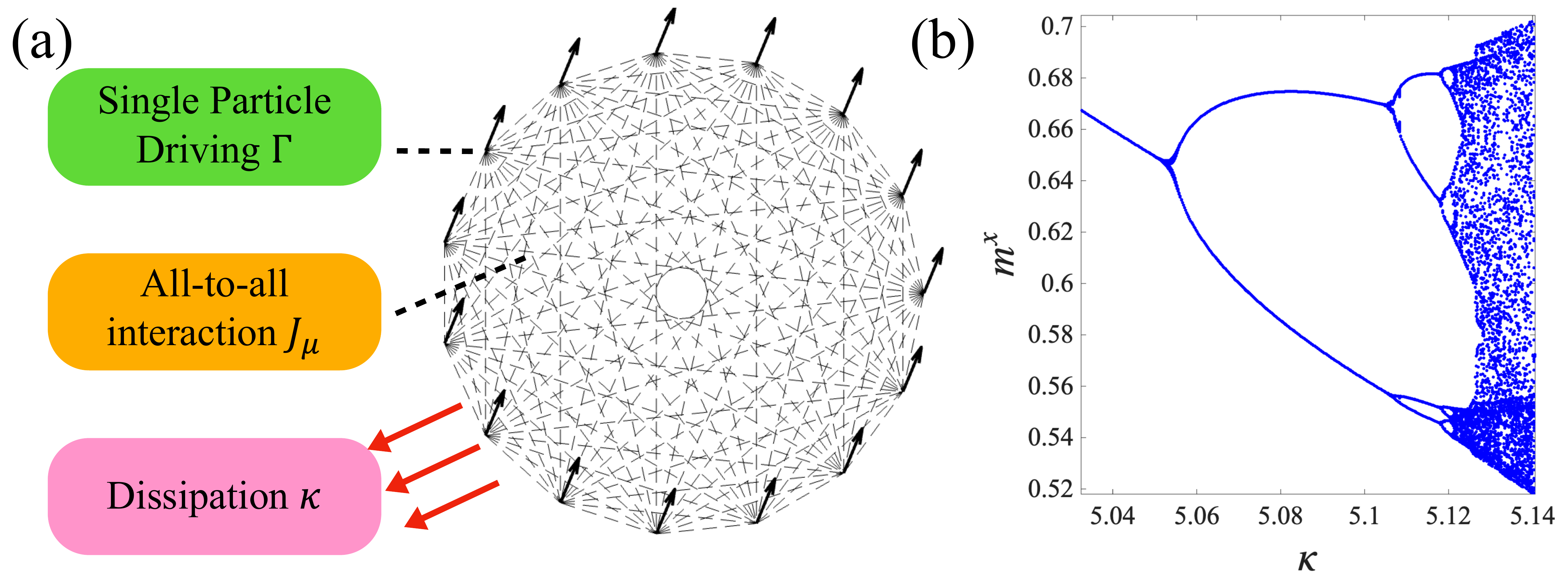}
\caption{(a) A schematic diagram of the spin ring with periodic single-particle driving, all-to-all interactions, and dissipation. (b) An example of a bifurcation process  from periodic behavior to chaos in the system.}
\label{fig1}
\end{figure}

Below, we investigate a periodically driven macrospin ensemble consisting of $S=\frac{1}{2}$ spins with anisotropic long-range all-to-all interactions and collective dissipation, governed by a Lindblad master equation. 
As a model, we use the following Hamiltonian~\cite{fu2022quantum, PhysRevResearch.5.L032024},  
\begin{align}
    \label{manybodyHamiltonian}
    H(t) =& \sum_{j=1}^N \frac{\Gamma}{2}\bigg(\sin{(\omega t)} \hat{\sigma}_j^x + [1-\cos{(\omega t)}] \hat{\sigma}_j^y \bigg)\nonumber\\
    &+ \sum_{\mu=x,y,z}\frac{J_{\mu}}{N} \sum_{i\neq j} \hat{\sigma}_i^{\mu} \hat{\sigma}_j^{\mu}.
\end{align}
Its first term describes the drive with an apparent period of $T=\frac{2{\pi}}{\omega}$, but, in fact, it accounts for a rotating (a frequency $\omega/2$) magnetic field whose magnitude also oscillates with the same frequency $\omega/2$ (and amplitude $\frac{\Gamma}{2}$). Here, $\hat{\sigma}^{\mu}_j\ (\mu=x,y,z)$ are the single-particle Pauli operators describing spins on each $j$-th site. For convenience, we use units of $\hbar = 1$ and measure time in units of $2\pi/\omega$ (implemented by using $\omega=2\pi$, so that the driving period is $1$).
The second term in $H$ introduces a uniform and anisotropic all-to-all pairwise interaction. 
Here, ``uniform" means that $J_x$, $J_y$, and $J_z$ are constants across all sites, and ``anisotropic" means $J_x$, $J_y$, and $J_z$ are not all equal. For other details, see Appendix~\ref{secA1}. 

We analyze this system in both the ``classical limit" and the ``quantum case". The ``classical limit" is implemented as the analysis of a ``macrospin" corresponding to the mean-field description of the thermodynamic limit, $N\to\infty$ (it has already been established rigorously~\cite{PhysRevLett.133.150401} that, for open quantum many-body systems with infinite-range interactions, the mean-field description of macroscopic observables becomes exact for $N \to \infty$).  The 'quantum case' corresponds to the finite-$N$ system and evolution of the multi-spin systems upon the action of Hamjiltonian in Eq. (1). In the classical limit, we find a wealth of dynamical regimes, illustrated in Fig.~\ref{fig1}(b) and listed in Table~\ref{table1}: those include periodic oscillation, bifurcations, quasiperiodic behaviour, and transition to full chaos, all caused by the interplay between periodic drive, long-range interactions, and dissipation. In particular, we detect the transition to chaos with the help of the maximal Lyapunov exponent ~\cite{torres2024floquetlyapunovtheorynonautonomouslinear,alligood1998chaos,PhysRevE.103.052212,wang2023modeling,PhysRevA.109.013328,carpintero2020lyapunov,PhysRevB.109.064309,RevModPhys.57.617,ermakov2025periodicclassicaltrajectoriesquantum} (MLE), $\lambda_{\max}$, as a diagnostic tool which enables us to distinguish between unstable and stable classical dynamics. 
In the finite-$N$ quantum case, we simulate quantum dynamics of the system using the permutationally-invariant basis~\cite{sciolla2011dynamical,kerzner2025simplewayspreparingqudit,ballesteros2025deformationssymmetricsubspacequbit}, which is exact for several initial states selected and addressed in the reported modeling. 


\begin{figure*}[t]
\centering
\includegraphics[width=1\textwidth]{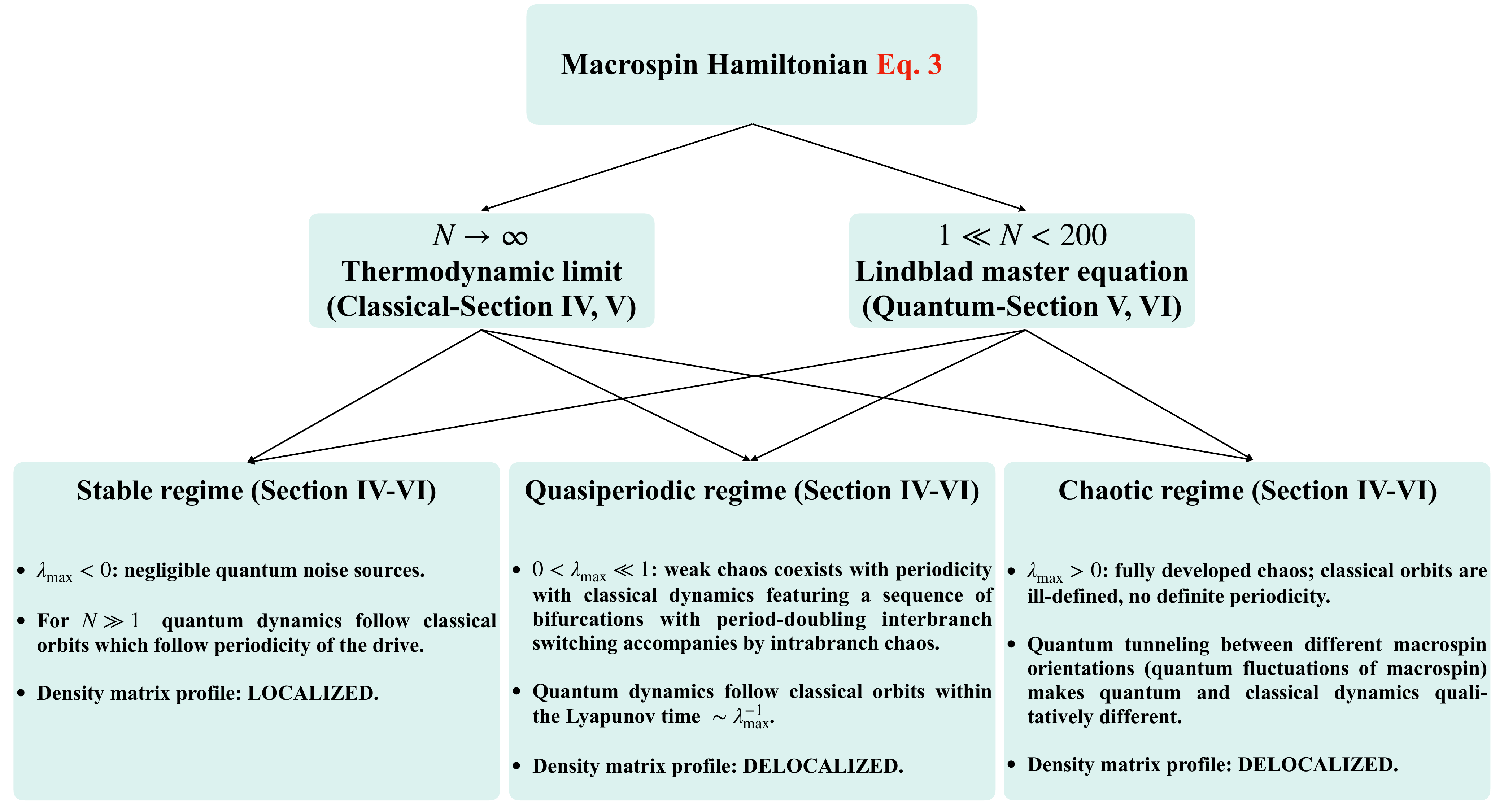}
\caption{Schematic overview of this work. The macrospin Hamiltonian [Eq.~\ref{collectiveHamiltonian}] branches into the classical thermodynamic limit ($N\to\infty$; Sec.~\ref{sec_bifurctaion}-\ref{sec_finiteNandinfiniteN}) and the quantum description via the Lindblad master equation ($1 \ll N < 200$; Sec.~\ref{sec_finiteNandinfiniteN}-\ref{Sec_densitymatrix}). Three dynamical regimes are identified according to the maximal Lyapunov exponent $\lambda_{\max}$: stable ($\lambda_{\max}<0$), quasiperiodic ($0<\lambda_{\max}\ll 1$), and chaotic ($\lambda_{\max}>0$). Key characteristics include the degree of quantum noise, agreement between quantum and classical trajectories, and localization of the density matrix.}
\label{logicFlow}
\end{figure*}

Based on the performed analysis, we note that, at the short timescales (up to the Lyapunov time $t_L \sim 1/\lambda_{\max}$), the evolutions of the quantum and classical systems closely agree, regardless of system size or the sign of the maximal Lyapunov exponent $\lambda_{\max}$. However, beyond the Lyapunov time, at $t>t_L$, such agreement is held no more, due to the tunneling between macrospin states in finite-$N$ systems that cause quantum fluctuations. That is, when the classical system exhibits chaotic dynamics and $\lambda_{\max} > 0$, chaos amplifies the effects of quantum fluctuations, leading to qualitative differences between the observables in a large-N quantum system and dynamics of their classical analog $\lambda_{\max} > 0$. This outcome agrees with the results of the recent studies \cite{PhysRevA.109.013328,PhysRevLett.133.150401} performed on a dissipative periodically driven system with all-to-all interactions with different drive and interaction anisotropy. In contrast, when $\lambda_{\max} < 0$, suggesting stable classical dynamics, quantum evolution of a system with for $N\gg 1$ can be expected to converge to its classical counterpart. However, we find that, whenever classical dynamics features bifurcations into period-2 orbits (consistent with the Feigenbaum constant~\cite{layek2015introduction}, or period-4, detected by the Fourier analysis of the time-dependent macrospin components), quantum fluctuations due to macrospin tunneling suppress longer-period cycles, causing quantum dynamics to retain simple period-1 oscillations. Thus, even in the stable regime, finite-N quantum effects qualitatively alter the long-time behavior and deviate dynamical evolution of quantum systems from classical dynamics, though the system does exhibit agreement between quantum and classical dynamics within the Lyapunov time during the initial stage of evolution. Additionally, we identify apparent discontinuities in the bifurcation diagrams that are not associated with true phase transitions, but instead arise from topological changes in the regions of attractors, in some parametric ranges forming fractal boundaries and self-similar structures, further highlighting the complexity of the underlying nonlinear dynamics.

\begin{table}
\caption{\label{table1}Representative examples illustrating dynamical regimes identified and discussed in this work. In the table, $\overline{0.0000}$ indicates that the MLE has an upper bound of 0.0000 at this parameter point: as the evolution time increases, the MLE approaches zero from negative values but never exceeds zero. This suggests the stable behavior.}
\begin{ruledtabular}
\begin{tabular}{cccccc}
 $\Gamma$ & $(J_x, J_y, J_z)$ & $\kappa$ & $\lambda_{\rm max}$
& Periodicity & Classification \\ \hline
0.078 & $(0, 1, 0)$ & $3.00$ & $\overline{0.0000}$ & 1 & stable \\
3.306 & $(0, 1, 0)$ & $0.02$ & $0.4827$ & none & chaotic \\
3.306 & $(0, 1, 0)$ & $3.00$ & $\overline{0.0000}$ & 1 & stable \\
3.306 & $(0, 1, 0)$ & $5.00$ & $\overline{0.0000}$ & 1 & stable \\
8.427 & $(0, 1, 0)$ & $1.00$ & $0.0284$ & 2 (not exact) & quasiperiodic \\
8.427 & $(0, 1, 0)$ & $3.00$ & $-0.0009$ & 2 & stable \\
\end{tabular}
\end{ruledtabular}
\end{table}

{The overall structure of this work is outlined in Fig.~\ref{logicFlow}. The analysis starts from the macrospin Hamiltonian [Eq.~\ref{collectiveHamiltonian}] and follows two main paths: the classical thermodynamic limit ($N\to\infty$), presented in Sec.~\ref{sec_bifurctaion} and Sec.~\ref{sec_finiteNandinfiniteN}, and the quantum evolution described by the Lindblad master equation for finite large system sizes ($1 \ll N < 200$), discussed in Sections Sec.~\ref{sec_finiteNandinfiniteN} and Sec.~\ref{Sec_densitymatrix}. The dynamics are classified into three regimes—stable, quasiperiodic, and chaotic—according to the maximal Lyapunov exponent. The diagram highlights the key features of classical and quantum behavior in each regime, including the degree of correspondence between classical and quantum trajectories, the influence of quantum fluctuations, and the localization properties of the density matrix.}

The manuscript below is organized as follows. In Sec.~\ref{sec_methodology}, we introduce the computational strategies for both the quantum case and the classical case. In Sec.~\ref{sec_LE}, we  describe the analysis of the maximal Lyapunov exponent (MLE). In Sec.~\ref{sec_bifurctaion}, we discuss bifurcations detected in the classical marcospin dynamics. We compare classical and quantum dynamics in Sec.~\ref{sec_finiteNandinfiniteN}, with the details of the density matrix profile computed for the quantum evolution presented in Sec.~\ref{Sec_densitymatrix}, and the overall finding summarised in Sec.\ref{sec_conclusion}.

\section{Methodology}\label{sec_methodology}
To investigate the dynamics of this periodically-driven dissipative system, we seek to compare the dynamical behavior exhibited in the quantum case with that in the thermodynamic limit ($N\to\infty$). In this section, we present the numerical approaches used to simulate the system's evolution in both cases. Specifically, we provide the explicit forms of the Lindblad master equation in both the quantum and classical limits, together with the form of the averaged magnetization operator $\mathbf{m}(t)$.

\subsection{Lindblad master equation}
Note that in the Hamiltonian given by Eq.~\eqref{manybodyHamiltonian}, the interaction strengths are rescaled by the system size with a factor $\frac{1}{N}$ to obtain a well-defined thermodynamic limit as $N\to\infty$~\cite{sciolla2011dynamical}. 
To study the macrospin dynamics, we further introduce the polarization operators 
\begin{align}
    \hat{S}^{\mu} = \frac{1}{N}\sum_i \hat{\sigma}^{\mu}_i, \quad (\mu=x,y,z)
\end{align}
where $[\hat{S}^{\alpha}, \hat{S}^{\beta}] = i\frac{2}{N}\epsilon_{\alpha\beta\gamma}\hat{S}^{\gamma}$ with ${\epsilon_{\alpha\beta\gamma}}$ denoting the Levi-Civita symbol. While different components of quantum macrospin do not commute and cannot be measured exactly simultaneously, in the limit of $N\to\infty$, quantum dynamics could be expected to coincide with classical. Using macrospin description, the Hamiltonian in Eq.~\eqref{manybodyHamiltonian} can be recast in the form of
\begin{align}\label{collectiveHamiltonian}
    H(t) =& \frac{N\Gamma}{2}\sin{(\omega t)}\hat{S}^x + \frac{N\Gamma}{2}[1-\cos{(\omega t)}]\hat{S}^y\nonumber\\
    &+ NJ_x (\hat{S}^x)^2 + NJ_y (\hat{S}^y)^2 + NJ_z (\hat{S}^z)^2,
\end{align}
where we ignore a constant by applying a global shift to the energy spectrum.

Furthermore, we introduce dissipation via a Lindblad master equation~\cite{manzano2020short}. 
In particular, we consider a collective decay process in which all spins dissipate coherently through a common reservoir. 
The jump operator drives the spins to align along the $-z$ direction and is given by
\begin{align}\label{jumpoperator}
    \hat{\sigma}^{-} &= \sum_j\hat{\sigma}_j^{-}= \frac{1}{2}\sum_j(\hat{\sigma}_j^x - i\hat{\sigma}_j^y) = \frac{N}{2}(\hat{S}^x - i\hat{S}^y).
\end{align}
Such a jump operator gives rise to the Lindblad master equation 
\begin{align}\label{LindbladEq}
    \frac{d\hat{\rho}(t)}{dt} = -i[H(t),\hat{\rho}(t)] + \frac{\kappa}{N}[2\hat{\sigma}^{-}\hat{\rho}(t)\hat{\sigma}^{+}-\{\hat{\rho}(t), \hat{\sigma}^{+}\hat{\sigma}^{-}\}].
\end{align}
Here $\kappa>0$ is the dissipation strength. 
We use $\hat{\rho}(t)$ to denote the density matrix, and a factor $\frac{1}{N}$ in the dissipation term ensures that the dissipative contribution remains meaningful in the thermodynamic limit.

We focus on the evolution of the averaged magnetization $\mathbf{m}(t) = \langle \hat{\mathbf{S}}(t)\rangle$, which is the expectation value of the polarization operators defined earlier. The expectation value of an observable $\hat{\mathcal{O}}$ at time $t$ is defined by $\langle\hat{\mathcal{O}}(t)\rangle = \mathbf{Tr}[\hat{\rho}(t)\cdot\hat{\mathcal{O}}]$.

\subsection{Finite-$N$ quantum evolution}
For the sake of simplicity in our analysis, and also with potential future experimental implementations in mind, we consistently adopt as the initial state of our many-body system a product state composed of identical single-particle states across all sites. 
That is, the many-body initial state is chosen to be a product state with all the single-particle states being the same. 
Thus, the many-body state is invariant under any permutation of the sites $(1,2,\dots,N)\to\sigma(1,2,\dots, N)$, where $\sigma$ denotes an element of the permutation group. 
It is worth noting that the Lindblad master equation~\eqref{LindbladEq} also respects this permutational invariance. 
Therefore, we expect that this symmetry is preserved throughout the system’s evolution~\cite{PhysRevA.89.022118}.

We start by considering a quantum system with $N$ spin-$\frac{1}{2}$ particles. 
The Hilbert space of such a system grows exponentially with system size, having a dimension of $2^N$. 
As a result, direct calculations within the full Hilbert space become computationally difficult. Fortunately, the system in Eqs.~\eqref{manybodyHamiltonian}-\eqref{LindbladEq} possesses the permutation symmetry, which allows us to significantly reduce the dimension of the Hilbert space to $N+1$ by introducing the so-called Dicke basis with maximum angular momentum~\cite{sciolla2011dynamical}:
\begin{align}\label{Dickebasis}
        \ket{M} = \sqrt{\frac{1}{{{\rm C}_{N}^{N/2 + M}}}}\sum_{\sum_j s_j^z = M} \ket{s_1^z, s_2^z, \cdots, s_N^z},
\end{align}
where ${\rm C}_n^m=\frac{n!}{m!(n-m)!}$ is the binomial coefficient. We have $s_j^z = \pm\frac{1}{2}$, implying whether the $j$-th spin is parallel or anti-parallel to the $z$-direction. We use the total magnetization $M$ in the $z$-direction to label the basis vectors and $M\in\{-\frac{N}{2}, -\frac{N}{2}+1, \cdots, \frac{N}{2}-1, \frac{N}{2}\}$. This reduces a $2^N$-dimensional exponentially large Hilbert space to a $(N+1)$-dimensional subspace spanned by $\{\ket{M}\}$, and the density matrix is then restricted to the $(N+1)\times(N+1)$-dimensional space spanned by $\{\ketbra{M}{M'}\}$, given by 
\begin{align}
    \hat{\rho}(t) = \sum_{M, M'}\rho_{M, M'}(t)\ketbra{M}{M'}. 
\end{align}

In addition, we utilize the following creation and annihilation relations: 
\begin{align}\label{crea_anni}
  \hat{S}^x\ket{M}&=\mathcal{F}_M\ket{M-1} + \mathcal{F}_{M+1}\ket{M+1}, \nonumber\\
  \hat{S}^y\ket{M}&=i\mathcal{F}_M\ket{M-1} - i\mathcal{F}_{M+1}\ket{M+1}, \nonumber\\
  \hat{S}^z\ket{M}&=\frac{2M}{N}\ket{M},
\end{align}
where $\mathcal{F}_{M}=\sqrt{(\frac{1}{2}+\frac{M}{N})(\frac{1}{2}-\frac{M}{N}+\frac{1}{N})}$ is a function of $M$.
Then we expand the Lindblad master equation~\eqref{LindbladEq} in this basis with relations in Eq.~\eqref{crea_anni}, obtain the evolution equations for the matrix elements of $\hat{\rho}(t)$ as follows: 
\begin{widetext}
\begin{align}\label{finiteNLindblad}
    &\frac{1}{N}\frac{d}{dt}\rho_{M, M'}(t) =\nonumber\\
    &\bigg\{-i\frac{\Gamma}{2}\sin{(\omega t)} - \frac{\Gamma}{2}[1-\cos{(\omega t)}]\bigg\}\mathcal{F}_{M}\rho_{M-1, M'}(t) + \bigg\{-i\frac{\Gamma}{2}\sin{(\omega t)} + \frac{\Gamma}{2}[1-\cos{(\omega t)}]\bigg\}\mathcal{F}_{M+1}\rho_{M+1, M'}(t)\nonumber\\
    &+ \bigg\{i\frac{\Gamma}{2}\sin{(\omega t)} -\frac{\Gamma}{2}[1-\cos{(\omega t)}]\bigg\}\mathcal{F}_{M'}\rho_{M, M'-1}(t) + \bigg\{i\frac{\Gamma}{2}\sin{(\omega t)} + \frac{\Gamma}{2}[1-\cos{(\omega t)}]\bigg\}\mathcal{F}_{M'+1}\rho_{M, M'+1}(t)\nonumber\\
    &+ \bigg\{-iJ_x+iJ_y\bigg\}\mathcal{F}_{M}\mathcal{F}_{M-1}\rho_{M-2, M'}(t) + \bigg\{-iJ_x+iJ_y\bigg\}\mathcal{F}_{M+2}\mathcal{F}_{M+1}\rho_{M+2, M'}(t)\nonumber\\
    &+ \bigg\{iJ_x-iJ_y\bigg\}\mathcal{F}_{M'}\mathcal{F}_{M'-1}\rho_{M, M'-2}(t) + \bigg\{iJ_x-iJ_y\bigg\}\mathcal{F}_{M'+2}\mathcal{F}_{M'+1}\rho_{M, M'+2}(t)\nonumber\\
    &+\bigg\{-iJ_z(\frac{M}{N})^2 +iJ_z(\frac{M'}{N})^2\bigg\}\rho_{M, M'}\nonumber\\
    &+2\kappa\mathcal{F}_{M+1}\mathcal{F}_{M'+1}\rho_{M+1, M'+1} - \kappa[(\mathcal{F}_{M})^2 + (\mathcal{F}_{M'})^2]\rho_{M, M'}.
\end{align}
\end{widetext}
 
As for the initial state, we consider a product state with all spins aligned in the same initial polarization direction set by the azimuthal angles $(\theta, \phi)$, with $\theta\in[0, \pi]$, $\phi\in[0,2\pi)$. In the Dicke basis in Eq.~\eqref{Dickebasis}, we have
\begin{align}\label{finiteNInitialstate}
    \ket{\theta, \phi}=\sum_{M=-\frac{N}{2}}^{\frac{N}{2}}(\cos{\frac{\theta}{2}})^{\frac{N}{2}+M}(e^{i\phi}\sin{\frac{\theta}{2}})^{\frac{N}{2}-M}\sqrt{C_N^{\frac{N}{2}+M}}\ket{M}.
\end{align}

\begin{figure*}[ht]
\centering
\includegraphics[width=0.92\textwidth]{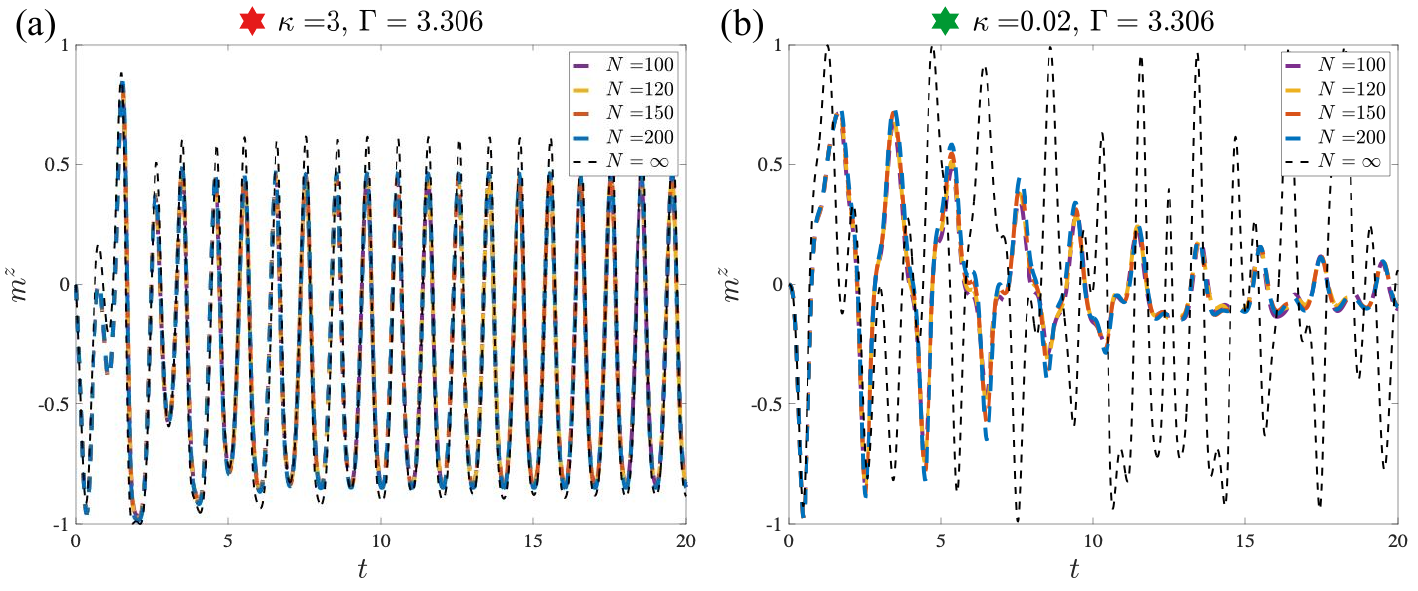}
\caption{Two representative examples of the evolution of $m^z(t)$ in finite-size quantum systems with system sizes ranging from 100 to 200, shown in different colors. The corresponding dissipation strength $\kappa$ and driving strength $\Gamma$ are indicated at the top of each panel. In addition, the parameter combinations corresponding to the two plots are also indicated in Fig.~\ref{MLE_PD} using the same symbols as in the titles. In both panels, the initial condition is set to be an $x$-polarized state. Black dashed lines represent the corresponding classical orbits in the thermodynamic limit.}
\label{example_finiteN}
\end{figure*}

We simulate the quantum evolution of the state in Eq.~\eqref{finiteNInitialstate} by numerically solving the differential equation for the density matrix given by Eq.~\eqref{finiteNLindblad}. 
This evolution is computed by the $4$-order Runge–Kutta (RK4) method (unless stated otherwise). Since we only need to update the density matrix elements $\rho_{M, M'}(t)$ at each time step of RK4 algorithm (of which there are $(N+1)^2$ in total), we are able to access large systems, with the number of spins up to $N\sim10^2-10^3$. See Appendix~\ref{secA2} for details and a test performed to demonstrate that the computed evolution is exactly restricted in the Dicke subspace defined in Eq.~\eqref{Dickebasis}.

With the computed $\hat{\rho}(t)$, we can analyze the evolution of the polarization vector $\mathbf{m}(t) = (m^x(t), m^y(t), m^z(t))$, whose components are given by
\begin{align}
    m^x(t) &= \sum_{M=-\frac{N}{2}}^{\frac{N}{2}} \mathcal{F}_{M+1}\rho_{M, M+1}(t) + \mathcal{F}_{M}\rho_{M, M-1}(t),\nonumber\\
    m^y(t) &= \sum_{M=-\frac{N}{2}}^{\frac{N}{2}} i\mathcal{F}_{M+1}\rho_{M, M+1}(t) - i\mathcal{F}_{M}\rho_{M, M-1}(t),\nonumber\\
    m^z(t) &= \sum_{M=-\frac{N}{2}}^{\frac{N}{2}} \frac{2M}{N}\rho_{M, M}(t).
\end{align}

Figure~\ref{example_finiteN} shows two representative examples of $m^z(t)$ evolutions for different parameter choices in Hamiltonian~\eqref{manybodyHamiltonian} and dissipation strength $\kappa$. For reference, classical evolution for the same observable (in the thermodynamic limit) is shown, too. 
We observe in both Fig.~\ref{example_finiteN}(a) and \ref{example_finiteN}(b) that the quantum-computed evolution appears to converge to a well-defined orbit as $N$ increases. Naturally, one could expect this limiting orbit to coincide with the classical dynamics results (as the thermodynamic limit). This expectation is indeed fulfilled in Fig.~\ref{example_finiteN}(a): both the quantum dynamics and the classical orbit exhibit periodic oscillations (due to the periodic driving), and as $N$ increases, the quantum orbits approach the classical one, as anticipated. However, the dynamics shown in Fig.~\ref{example_finiteN}(b) are markedly different: the convergent (upon increase of $N$) quantum orbits obtained from Eq.~\eqref{finiteNLindblad} and Eq.~\eqref{finiteNInitialstate} show no sign of approaching the classical orbit. In fact, the classical orbit itself appears highly irregular and lacks any apparent structure. 

This contrasting behavior in the two panels arises from the fact that the system resides in fundamentally different dynamical phases in the two cases: the dynamics in Fig.~\ref{example_finiteN}(a) is stable, whereas that in Fig.~\ref{example_finiteN}(b) is unstable. 
In Section~\ref{sec_LE}, we will employ a powerful quantitative diagnostic to characterize this dynamical stability versus instability using Lyapunov exponents analysis.

\subsection{Infinite-$N$ mean-field classical limit}
We adopt the same order parameters $\mathbf{m}(t) = \langle \hat{\mathbf{S}}(t)\rangle$, and employ a mean-field approach---commonly used to analyze models with infinite-range interactions~\cite{PhysRevA.103.013306,PhysRevA.109.013328,PhysRevA.40.6800,PhysRevA.98.063815,PhysRevB.104.014307,PhysRevLett.113.210401,PhysRevLett.119.190402, PhysRevLett.121.035301,PhysRevX.6.031011,PhysRevB.101.214302,PhysRevLett.110.257204}---to derive the set of classical equations governing the dynamics in the thermodynamic limit  ($N\to\infty$). 
In particular, it is often legitimate to assume for such models with infinite-range interactions that $\langle \hat{S}^{\alpha} \hat{S}^{\beta} \rangle = \langle \hat{S}^{\alpha} \rangle \langle \hat{S}^{\beta} \rangle$ as $N \to \infty$~\cite{PhysRevA.103.013306,PhysRevLett.133.150401}. 
By directly substituting the polarization operators into the Lindblad Eq.~\eqref{LindbladEq}, we obtain the following set of nonautonomous nonlinear differential equations~\cite{torres2024floquetlyapunovtheorynonautonomouslinear} that describes the time evolution of the order parameters (see Appendix~\ref{secA3} for its derivation):
\begin{align}\label{classicalEq}
    \frac{d}{dt} m^x =& \Gamma[1-\cos{(\omega t)}]m^z + 4(J_y-J_z)m^ym^z + \kappa m^x m^z,\nonumber\\
    \frac{d}{dt} m^y =& -\Gamma\sin{(\omega t)}m^z + 4(J_z-J_x)m^zm^x + \kappa m^y m^z,\nonumber\\
    \frac{d}{dt} m^z =& \Gamma\sin{(\omega t)}m^y - \Gamma[1-\cos{(\omega t)}]m^x \\
    &+ 4(J_x-J_y)m^xm^y - \kappa \bigg((m^x)^2+(m^y)^2\bigg),
    \nonumber
\end{align}
with $|{\mathbf{m}}(t)|=\sqrt{(m^x)^2+(m^y)^2+(m^z)^2}$ being a conserved quantity. 
In fact, Eq.~\eqref{classicalEq} describes a dynamical system evolving on a sphere, where each point stands for different Dicke states. 
In particular, we can initialize the system such that all spins point to the positive $x$-direction by setting $\mathbf{m}(0)=(1,0,0)$. The evolution in the thermodynamic limit is also calculated by the RK4 method.

\section{The Lyapunov Analysis}\label{sec_LE}
\begin{figure*}[ht]
\centering
\includegraphics[width=1\textwidth]{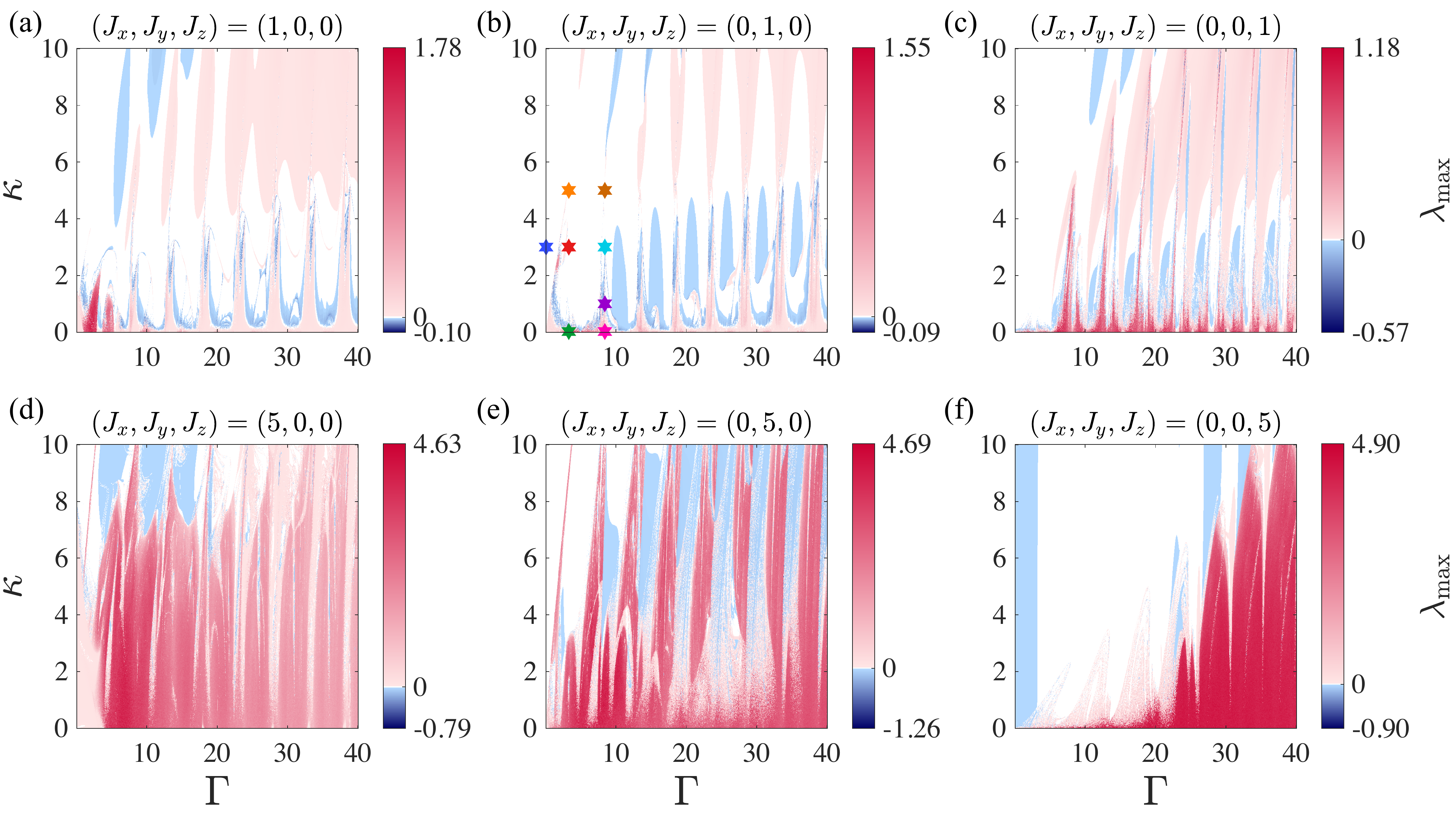}
\caption{\label{MLE_PD} (Classical) The Phase diagrams of the MLE for different choices of the interaction strength as a function of the driving strength $\Gamma$ and the dissipation strength $\kappa$. The initial condition is uniformly set to be $\mathbf{m}(0)=(1,0,0)$. We use red to indicate the chaotic phase where the MLE is positive, and white to denote regions where the MLE is zero. Blue is used to mark the stable phase with negative MLE values, allowing for a clear visual distinction between regions with positive and negative MLE. The interaction strength is indicated above each panel. Those parameter combinations to be analyzed later are marked by colored stars in (b).
}
\end{figure*}

Equations~\eqref{finiteNLindblad} and~\eqref{classicalEq} enable us to analyze the system dynamics in two complementary regimes: the quantum evolution for a finite system size and the classical evolution in the thermodynamic limit, respectively. For any chosen initial state, we refer to its phase space trajectory as an orbit. 
The instability observed in Fig.~\ref{example_finiteN}(b) is a manifestation of chaotic dynamics, identifiable via its orbit structure. 
In nonlinear systems, chaos typically arises from the sensitivity of orbits to infinitesimal perturbations in their initial conditions, quantified using the Lyapunov exponent. To quantitatively diagnose the emergence of chaos, we analyze the spectrum of Lyapunov exponents, detecting the \textbf{maximal Lyapunov exponent (MLE)}, and analyze its dependence on input parameters. The MLE serves as a rigorous measure of the rate at which nearby orbits in phase space diverge, providing a powerful tool for analyzing dynamical instabilities. 

Therefore, for an orbit $\boldsymbol{\mathcal{S}}(t)$, we introduce an additional perturbation to the initial condition, $\boldsymbol{\mathcal{S}}_1(0) = \boldsymbol{\mathcal{S}}(0) + \boldsymbol{\delta}(0)$, where $\boldsymbol{\delta}(0)$ is very small, and we can obtain a new orbit $\boldsymbol{\mathcal{S}}_1(t) = \boldsymbol{\mathcal{S}}(t) + \boldsymbol{\delta}(t)$. Then the MLE of $\boldsymbol{\mathcal{S}}(t)$ can be formally quantified by
\begin{align}\label{LE_def}
    \lambda_{\mathrm{max}} = \underset{t\longrightarrow\infty}{\lim}\underset{|\boldsymbol{\delta}(0)|\longrightarrow 0}{\lim}\frac{1}{t}\ln\frac{\boldsymbol{\delta}(t)}{\boldsymbol{\delta}(0)}.
\end{align}
If $\lambda_{\mathrm{max}}>0$, any initial perturbation grows exponentially in time, indicating that $\boldsymbol{\mathcal{S}}(t)$ is extremely sensitive to its initial condition. 
In this regime, the system exhibits chaotic dynamics, and its long-time evolution becomes effectively unpredictable. 
Conversely, if $\lambda_{\mathrm{max}}<0$, any finite initial perturbation decays exponentially in time, so nearby trajectories contract and approach a common, stable orbit or fixed point—signaling a dynamically stable regime. Finally, $\lambda_{\mathrm{max}}=0$ corresponds to a marginal case at the boundary between stability and chaos.

In the thermodynamic limit, the classical equation Eq.~\eqref{classicalEq} describes a dynamical system whose state evolves on a sphere. This geometric structure naturally allows us to use the MLE to characterize dynamical stability and detect chaos. 
To compute the Lyapunov spectrum with high accuracy, we employ a standard algorithm based on QR decomposition and orthogonalization~\cite{geist1990comparison,benettin1980lyapunov}. Specifically, the tangent-space vectors are evolved together with the system’s phase space orbit, and after each finite time step, they are re-orthogonalized using QR decomposition. The logarithms of the norms of the resulting orthogonal factors are accumulated and averaged over long times, yielding all Lyapunov exponents in a stable and efficient manner. The largest exponent in this spectrum is identified as the MLE. Figure~\ref{MLE_PD} presents the phase diagrams of the MLE for different interaction strengths, where the initial states are uniformly chosen to be $x$-polarized, i.e., $\mathbf{m}(0)=(1,0,0)$. 

In the quantum case with a finite system size, however, the situation becomes substantially different. 
While Lyapunov exponents can be rigorously defined in the thermodynamic limit, they are often employed as effective descriptors in finite-size systems. 
Specifically, the MLE computed in the thermodynamic limit quantifies the average rate of divergence or convergence of nearby orbits, characterizing the intrinsic stability nature of the dynamics.

In contrast, true Lyapunov exponents cannot be strictly defined for finite-size systems; nevertheless, their early-time evolution frequently exhibits transient exponential sensitivity to initial perturbations. 
This enables the well-defined MLE in the thermodynamic limit to serve as a useful tool for interpreting short- to intermediate-time dynamics in finite-size systems~\cite{PhysRevLett.133.150401}. 
Caution is required, however, when extrapolating this interpretation to longer times, as finite-size effects—such as the saturation of perturbation growth—can substantially affect the system’s behavior. 
Moreover, in open quantum systems governed by Lindblad dynamics, dissipation tends to drive the system toward steady states, further constraining the Lyapunov-type growth at late times. 
Therefore, while the MLE obtained from the thermodynamic limit offers valuable insights, it should be regarded as an effective indicator rather than a strictly valid measure in finite systems.

In the next section, we investigate the dynamics in the thermodynamic limit and utilize the MLE obtained from this limit to qualitatively characterize the system's dynamical behavior.

\section{Transition to chaos: Bifurcation in classical dynamics}\label{sec_bifurctaion}
\begin{figure*}[t]
\centering
\includegraphics[width=1\textwidth]{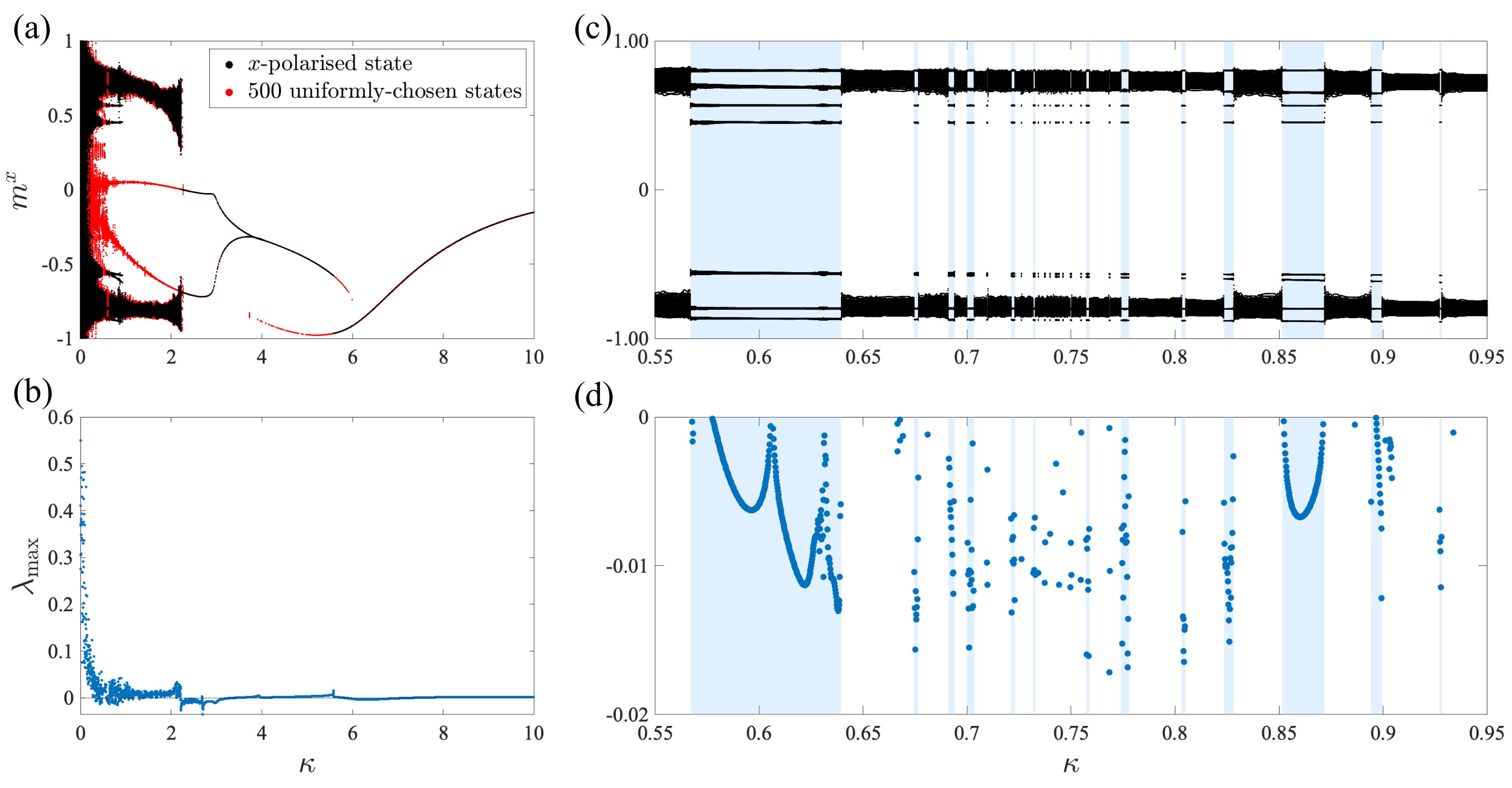}
\caption{\label{010bifurcationLinecut}
(Classical) (a) \textbf{Bifurcation diagram} of $m^x(t)$ for driving strength $\Gamma=8.427$. Black dots correspond to dynamics starting from the $x$-polarized initial state, highlighting state-specific behavior. Red dots represent results from $500$ uniformly sampled initial states over the sphere, capturing the global dynamical structure. (b) MLE as a function of $\kappa$, computed for the $x$-polarized initial state. (c, d) Zoomed-in views of selected sub-regions of panels (a) and (b), respectively. Blue-shaded areas mark stable windows, identified by both stable periodic orbits and negative MLE values.}
\end{figure*}

From a mathematical perspective, it is well established that rich dynamical phenomena such as bifurcations~\cite{RevModPhys.63.991} often arise near regions where the MLE changes sign, i.e., transitions between positive and negative values. 
As illustrated in Fig.~\ref{MLE_PD}, the MLE phase diagram displays qualitatively distinct structures for different interaction strengths, implying that tuning the interaction strength can dramatically modify the system’s dynamical behavior. 
Since bifurcations are a well-known route to chaos, these observations suggest that the system can undergo a variety of nonlinear dynamical transitions. In the following, we investigate the rich dynamical behaviors exhibited by the system in the thermodynamic limit.

We first focus on the case with an interaction strength $(J_x,J_y,J_z) = (0,1,0)$, whose MLE phase diagram is shown in Fig.~\ref{MLE_PD}(b). 
To resolve the bifurcations and distinguish different dynamical behaviors (periodic, quasiperiodic, and chaotic), we solve the classical equations of motion~\eqref{classicalEq} from various initial conditions and construct a stroboscopic bifurcation diagram. 
Concretely, for each set of parameters, we first discard an initial transient of $200$ driving periods, and then sample $m^x(t)$ at $1000$ stroboscopic times,
\[
t = 201,202,\cdots,1200,
\]
i.e., once per driving period (note that we measure time in the units of the fundamental period, $2\pi/\omega$). 
These stroboscopic values capture the long-time attractor of the dynamics and its periodicity. 
Figure~\ref{010bifurcationLinecut}(a) shows the resulting bifurcation diagram of $m^x(t)$ as a function of $\kappa$.
Here, we fix the driving strength at $\Gamma = 8.427$.
As for the initial condition, we adopt two strategies marked by different colors:
\begin{enumerate}
    \item \textit{Black:} We consider a fixed $x$-polarized state as the initial condition for all evolutions, allowing for direct comparison with Fig.~\ref{MLE_PD}, which is based on the same choice.
    \item \textit{Red:} For a better understanding, we use all the points over the two-dimensional sphere as the initial states. In practice, we uniformly sample $500$ initial states from the sphere. We then compute the \textbf{bifurcation diagram} for each of these initial states and plot them together in Fig.~\ref{010bifurcationLinecut} .
\end{enumerate}

It is evident from Fig.~\ref{010bifurcationLinecut}(a) that the set of black points is a subset of the red points. This is to be expected, as the black points correspond to orbits originating specifically from the $x$-polarized initial state, and thus only capture the attractors reached from that particular initial condition under different parameter settings. In contrast, the red points reflect the global dynamical landscape of the system, representing all possible stable states  accessible from the whole range of initial conditions. A more in-depth discussion of the attractors observed here will be presented later in this section. 

Fig.~\ref{010bifurcationLinecut}(b) is a linecut of the MLE phase diagram in Fig.~\ref{MLE_PD}(b) along $\Gamma=8.427$, in which we only adopt the $x$-polarized state as the initial condition for comparison with those black dots in Fig.~\ref{010bifurcationLinecut}(a). 
We observe that in the region around $\kappa<0.1$, the MLE takes significantly positive values, indicating that the system resides in an unstable, chaotic phase. This is also evident in Fig.~\ref{010bifurcationLinecut}(a), where the values of $m^x(t)$, within this parameter range, appear to densely explore the entire state space. As $\kappa$ increases slightly, the MLE quickly approaches zero, remaining small but positive. This suggests that while the system remains in an unstable phase, the chaotic behavior becomes less pronounced. 
In fact, the system enters a quasiperiodic regime. For instance, when $\kappa=1$, the system’s classical dynamics---shown by the dashed line in Fig.~\ref{finiteN_dynamics}(d) below---exhibit a quasi-2-periodic pattern that is not strictly exact. This occurs because, with increasing dissipation strength, the system begins to develop a more stable response. However, the small but still positive MLE tends to destabilize this process and pull the system back towards chaotic behavior. The resulting dynamics reflect a compromise between these two competing tendencies. As shown in Fig.~\ref{010bifurcationLinecut}(a), within this parameter regime, the observable $m^x(t)$ repeatedly jumps between two branches in an attempt to establish period-$2$ behavior. However, each return to a given branch occurs at a different and highly unpredictable position, indicating the presence of chaos within each branch. The interplay between the chaotic intrabranch motion and the periodic interbranch switching gives rise to the quasiperiodicity observed in this regime. 

An interesting observation arises in the parameter range $0.5<\kappa<1$, where the MLE linecut calculation in Fig.~\ref{010bifurcationLinecut}(b) exhibits many negative values. Initially, we suspected these might be artefacts of numerical inaccuracies in the MLE algorithm or unpredictable features intrinsic to the quasiperiodic regime. However, a closer inspection of the bifurcation diagram in Fig.~\ref{010bifurcationLinecut}(a) reveals the presence of small windows within the same parameter range, in which the system’s dynamical behaviour appears qualitatively distinct from its neighbours---more regular and seemingly more periodic. 

To explore the latter point, we performed a more refined computation of both the bifurcation diagram and the MLE linecut, presented in Fig.~\ref{010bifurcationLinecut}(c) and Fig.~\ref{010bifurcationLinecut}(d), respectively. These negative MLE values indeed offer meaningful insights. 
As shown in Fig.~\ref{010bifurcationLinecut}(c), occasional small windows emerge within the broader quasiperiodic regime where the observable $m^x(t)$ exhibits markedly periodic or predictable behaviour. 
We manually shaded some of these prominent regions in blue and applied the same shading to the corresponding regions in the MLE plot. 
In Fig.~\ref{010bifurcationLinecut}(d), these blue regions align precisely with the intervals where negative MLE values are obtained. 
This confirms that the negative MLE faithfully captures the presence of locally stable, periodic windows, thereby reinforcing the precision and reliability of the MLE algorithm. As a point of reference, such sudden appearances of local stability within an otherwise unstable phase are also found in the well-known logistic map~\cite{chen2021logistic} in certain parameter ranges.

As $\kappa$ increases further, we observe from Fig.~\ref{010bifurcationLinecut}(a) that in the range $\kappa\in[2.2,3.8]$, the system exhibits a well-defined period-$2$ behavior—for example, at $\kappa=3$, as illustrated by the dashed curves in Fig.~\ref{finiteN_dynamics}(e) and \ref{finiteN_dynamics}(f). 
Furthermore, it is noteworthy that in this regime, the sets of black and red points in the bifurcation diagram coincide perfectly. 
This indicates that, for any initial state chosen on the sphere, the system’s orbit converges, after a short transient, to the same period-$2$ orbit. 
We refer to such a stable orbit as an attractor, and the collection of initial conditions in the state space that evolve towards this orbit and remain confined to it defines the region of attraction of the attractor. 
Specifically, in this regime, we have a period-two attractor with the region of attraction being the whole sphere.

As $\kappa$ increases beyond $3.8$, the system’s attractor undergoes another transition—from a period-$2$ orbit to a trivial period-$1$ orbit. Interestingly, in the range $3.8<\kappa<6$, the system possesses two distinct period-$1$ attractors. 
Depending on the initial condition, the system evolves toward one of two attractors: a subset of initial states on the sphere (including the $x$-polarized state) is attracted to the orbit represented by the black points in the bifurcation diagram, while initial states in the complementary region are drawn to the second attractor shown in red.
In other words, the sphere of initial states is now divided into two regions. Two notable features emerge in this regime. 
First, around $\kappa\approx 5.5$, the $x$-polarized initial state is suddenly attracted by the other attractor, leading to a discontinuous jump in the bifurcation diagram. 
It is also worth noting that in Fig.~\ref{010bifurcationLinecut}(b), the MLE exhibits a sudden positive spike at the corresponding parameter value, strongly suggesting a transition in the attractor that captures the $x$-polarized initial state. 
Second, at $\kappa\approx6$, one of the attractors abruptly disappears, suggesting the occurrence of a sudden dynamical transition. 
To better understand the meaning of this apparent discontinuity in the bifurcation diagram, we must examine how the regions of attractors evolve on the sphere.

\begin{figure*}[t]
\centering
\includegraphics[width=1\textwidth]{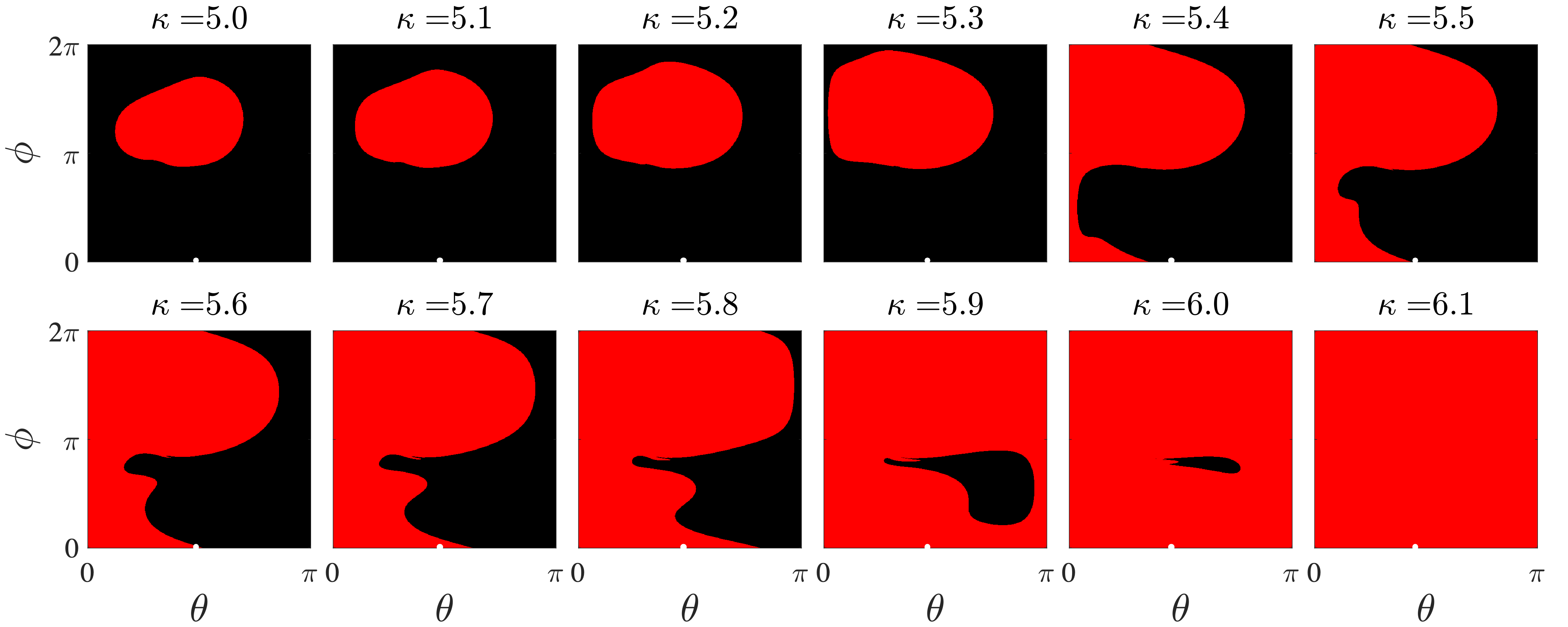}
\caption{\label{regionofattraction1}
(Classical) The evolution of the \textbf{regions of attractors} over the sphere as a function of $\kappa$. The value of $\kappa$ ranges from $5.0$ to $6.1$. We use the same black and red coloring scheme as in the corresponding parameter range of Fig.~\ref{010bifurcationLinecut}(a) to distinguish the regions of attractors of the two attractors. This allows for a direct comparison between the bifurcation structure and the corresponding regions of attractors. The white points marked in each panel correspond to the $x$-polarized initial state $(\frac{\pi}{2}, 0)$.
}
\end{figure*}

As shown in Fig.~\ref{regionofattraction1}, we unfold the sphere onto a plane to visualize the \textbf{regions of attractors}. Specifically, we use the coordinates $(\theta, \phi)$ to represent the spherical coordinates on the 2-dimensional sphere, with the mapping from the plane to the sphere (embedded in three-dimensional space) given by  
\begin{align}
    (\theta, \phi) \longrightarrow (\sin\theta \cos\phi, \sin\theta \sin\phi, \cos\theta).
\end{align}
It is important to note that this mapping is surjective but not injective. In particular, all points with $\theta = 0$ in the plane are mapped to the north pole of the sphere. By initializing the system at each point on the sphere and evolving it for a long time, we determine which attractor the orbit eventually converges to. Each point is then colored accordingly—red or black—depending on the corresponding attractor, resulting in Fig.~\ref{regionofattraction1}. 
We observe that, for $5.0<\kappa<5.3$, the boundaries between the regions of attractors deform slowly. 
However, starting from $\kappa = 5.4$, the boundaries begin to change rapidly. 
Note that the $x$-polarized state corresponds to the point $(\theta, \phi) = (\pi/2, 0)$ in this diagram. 
We further find that, as $\kappa$ increases from $5.5$ to $5.6$, the $x$-polarized state crosses the boundary between the two regions: at $\kappa = 5.5$, it lies within the black region, whereas at $\kappa = 5.6$, it has moved into the red region and remains there for all larger $\kappa$. This explains the abrupt jump observed in the bifurcation diagram of Fig.~\ref{010bifurcationLinecut}(a) near $\kappa \approx 5.5$: there is no actual discontinuity in the dynamics, but rather a smooth deformation of the regions causes the $x$-polarized state to switch attractors as the boundary passes over it. Furthermore, we note that when $\kappa$ exceeds $6.0$, the black region disappears entirely, and the system possesses a single global attractor. This is consistent with the bifurcation diagram in Fig.~\ref{010bifurcationLinecut}(a), where only one branch remains for $\kappa > 6.0$.
Additionally, we find regions of attractors at $\kappa\approx4$ which is discussed in detail in Appendix.\ref{secA4}.

In Fig.~\ref{sphereorbit_M_FFt}, we evolve the systems with $\Gamma = 8.427$ and different $\kappa$ for a total time of $200$ (with period set to be 1), and visualize the orbits over the time interval $[150, 200]$, shown in blue, across four distinct dynamical phases: chaotic, quasiperiodic, 2-periodic, and 1-periodic, respectively, over the spheres. 
To better examine the temporal structure, we also analyze the time evolution of the $m^x(t)$. 
Using Fourier transforms for both the \textbf{early-time transient regime} ($0$–$50$) and longer-time ($150$–$200$) signals. The corresponding Fourier spectra are shown in red and blue, respectively, to capture the periodicity or lack thereof.
As discussed previously, for $\kappa = 0.02$, the system resides in the chaotic phase. Even after discarding the transient behavior, the orbit shows no sign of convergence and instead densely explores the sphere, consistent with a lack of periodic structure. 

This is also reflected in the $m^x(t)$ evolution and its Fourier spectrum: neither the early-time nor the late-time signal displays any dominant frequency component, confirming the absence of periodicity.
At $\kappa = 1$, the system enters the quasiperiodic regime. The orbit begins to exhibit some coherence on the sphere, suggesting a weak tendency toward periodicity. The $m^x(t)$ evolution and its late-time Fourier spectrum reveal a prominent peak near the frequency corresponding to a 2-period, though additional smaller peaks are also present. These secondary peaks indicate that the system does not fully settle into an exact 2-periodic orbit, and thus the motion is classified as quasiperiodic.
When $\kappa=3$, the system reaches a fully stable phase characterized by a clean 2-periodic orbit. The trajectory on the sphere clearly converges to a well-defined loop, and the Fourier spectrum of the late-time signal shows a sharp, isolated peak at $f = 1/2$, with all other components vanishing—indicative of a pure 2-periodic motion. The early-time transient is still visible but quickly dies out.
Finally, for $\kappa = 5$, the increased dissipation further stabilizes the dynamics, leading to a 1-periodic regime. The corresponding Fourier spectra for late-time intervals show a dominant peak, confirming that the system reaches a unique, globally stable 1-periodic attractor.

\begin{figure*}[ht]
\centering
\includegraphics[width=0.9\textwidth]{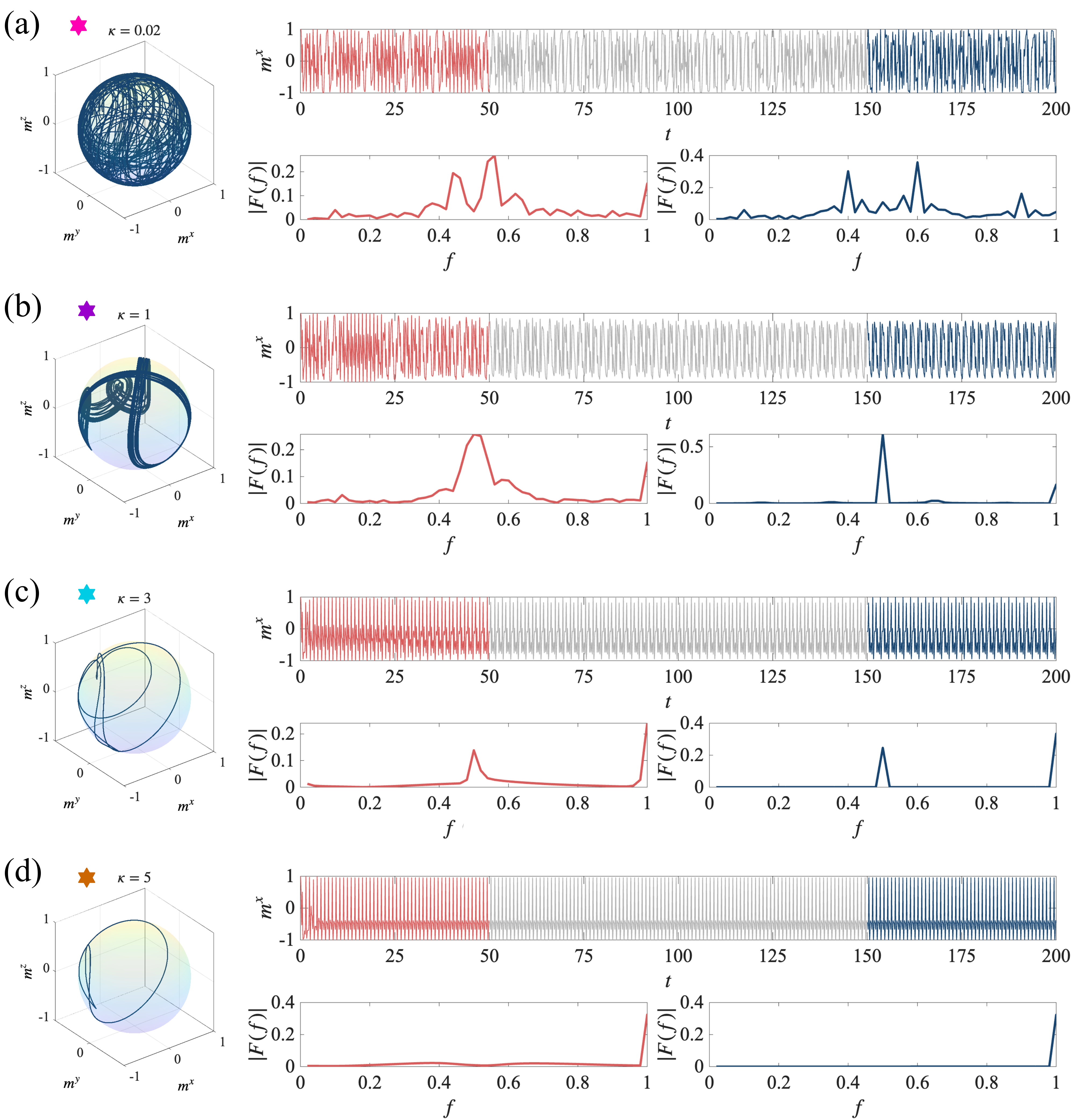}
\caption{\label{sphereorbit_M_FFt}
Classical dynamics of the system at four representative values of $\kappa$: (a) $\kappa = 0.02$ (chaotic), (b) $\kappa = 1$ (quasiperiodic), (c) $\kappa = 3$ (2-periodic phase), and (d) $\kappa = 5$ (1-periodic phase), all marked on Fig.~\ref{MLE_PD} using the same symbols, and with $\Gamma=8.427$. For each case, the system is evolved for a total time of $200$ periods. To distinguish between the \textbf{transient and steady-state dynamics}, we use red to denote the early-time window ($0<t<50$) and blue for the later-time window ($150<t<200$). For each $\kappa$, we also visualize the system’s orbit (during the interval $150<t<200$ interval) on the unit sphere to directly capture the qualitative nature of the motion, whereas for periodicity analysis, we select the $m^x(t)$ component and perform Fourier transforms of the data over both the early and late time intervals. Fourier analysis considers only frequencies corresponding to integer multiples of the driving period $T=2\pi/\omega=1$.}
\end{figure*}

\section{Quantum case \textit{vs} Classical case}\label{sec_finiteNandinfiniteN}
We are now in a position to compare the dynamics between quantum and classical cases. In regimes where the MLE of the classical system is negative, we expect the quantum orbits to converge towards those predicted by the classical equations in the thermodynamic limit, as the system size $N$ increases. 
In such stable regimes, quantum noise sources---such as tunneling and fluctuations---are rapidly suppressed by the negative MLE. 
Consequently, the dynamics settle onto a well-defined orbit that coincides with the stable orbit obtained in the thermodynamic limit.

The situation changes dramatically in regimes where the classical MLE is positive. 
Here, convergence between quantum and classical dynamics is generally not expected, even as $N$ grows. Two key factors are responsible for this discrepancy. 
First, a positive MLE causes any small perturbation to grow exponentially, leading to strong finite-size effects. 
Second, a positive MLE indicates that the classical orbit is chaotic, suggesting that the evolution of observables is highly sensitive to both initial conditions and perturbations incurred during time evolution. 
Even tiny deviations---either in the initial state or accumulated over time---undergo exponential amplification, resulting in substantial divergence of orbits. 
Moreover, numerical integration of Eq.~\eqref{finiteNLindblad} and Eq.~\eqref{classicalEq} introduces additional sources of instability. 
Although we adopt the relatively accurate RK4 method, it remains an approximation with errors of the order $\sim ({\Delta t})^4$. 
In chaotic regimes, the numerical errors inherent to any algorithm are also exponentially amplified by the positive MLE. 
This makes the notion of a unique ``true'' classical orbit fundamentally ill-defined: in such chaotic phases, the reference orbit itself is inherently unstable and unreliable. Thus, it is neither meaningful nor reasonable to expect the quantum dynamics to converge to such an ill-defined orbit.

Fortunately, there exists an early-time window before chaos fully manifests. 
Specifically, for the time window $t<t_L\sim1/\lambda_{\mathrm{max}}$ (often referred to as the Lyapunov time), the exponential amplification of small perturbations has not yet reached a macroscopic level. 
Within this short time, the influence of chaos is negligible, and the system behaves as if it were effectively stable. 
As a result, the classical orbit can be regarded as well-defined in this time interval, and quantum evolutions exhibit convergence toward the classical orbit as $N$ increases within this time scale~\cite{PhysRevLett.133.150401}.

\begin{figure*}[t]
\centering
\includegraphics[width=1\textwidth]{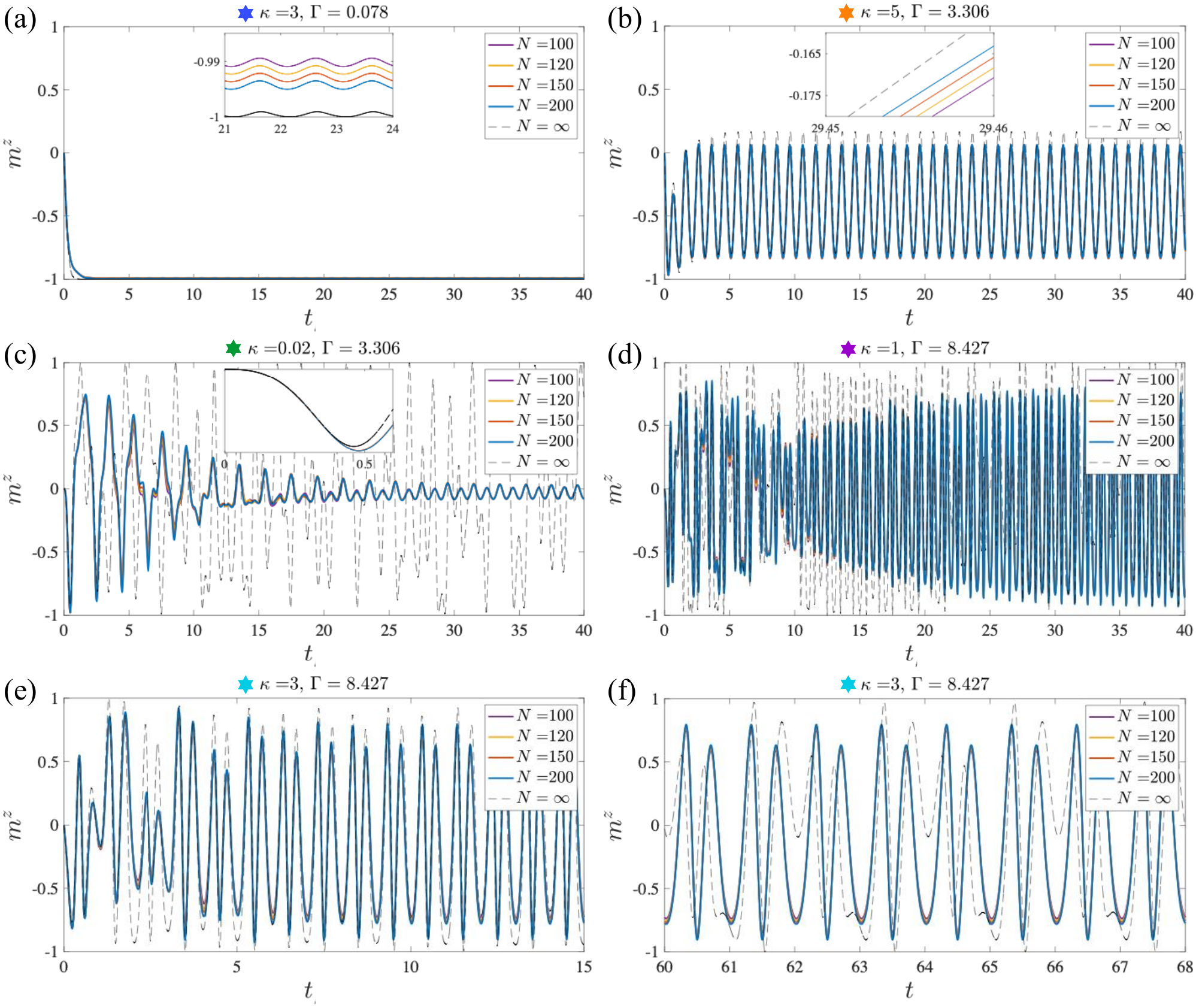}
\caption{\label{finiteN_dynamics}The evolution of $m^z(t)$ for both finite-size quantum system (colored line) and thermodynamic-limit system (dashed line) with different choices of parameters. (a) period-1: $\kappa=3$, $\Gamma=0.078$; (b) period-1: $\kappa=5$, $\Gamma=3.306$; (c) chaotic: $\kappa=0.02$, $\Gamma=3.306$; (d) quasiperiodic: $\kappa=1$, $\Gamma=8.427$; (e) and (f) period-2: $\kappa=3$, $\Gamma=8.427$, for early-time and late-time. In all panels, we choose the $x$-polarized state as the initial condition. The parameter combinations corresponding to all panels are also indicated in Fig.~\ref{MLE_PD} using the same symbols as in the titles.}
\end{figure*}

\begin{figure*}[t]
\centering
\includegraphics[width=0.8\textwidth]{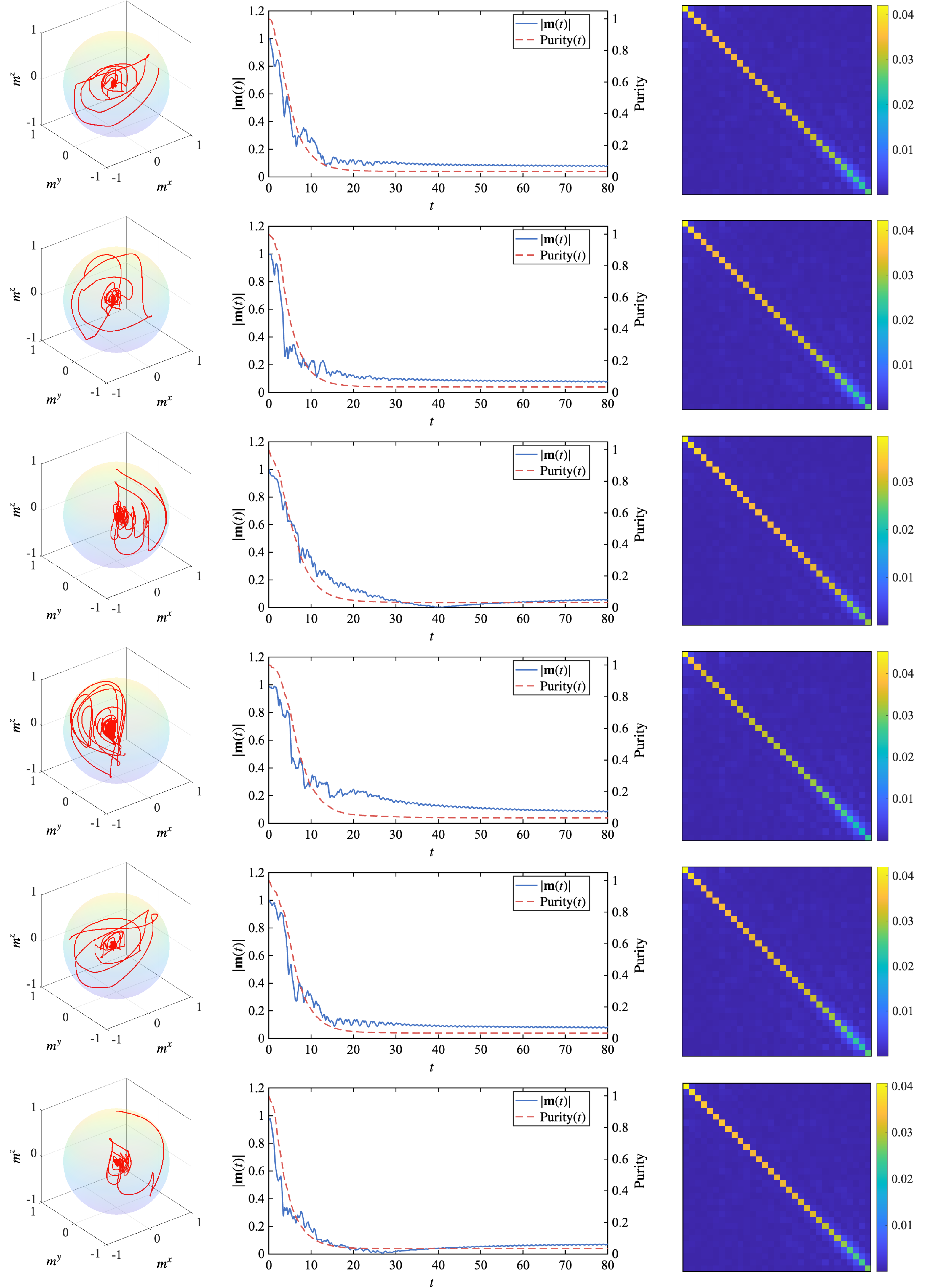}
\caption{\label{finiteN_onSphere} Quantum counterpart of the classical dynamics in the regime of chaos. Here, $\kappa = 0.02$ and $\Gamma = 3.306$. Six different initial states are chosen:
(a) $x$-polarized state: $\mathbf{m}(t) = (1, 0, 0)$;
(b) $y$-polarized state: $\mathbf{m}(t) = (0, 1, 0)$;
(c) $z$-polarized state: $\mathbf{m}(t) = (0, 0, 1)$;
(d) random state 1: $\mathbf{m}(t) = (0.4755, 0.3455, -0.8090)$;
(e) random state 2: $\mathbf{m}(t) = (-0.9045, 0.2939, 0.3090)$;
(f) random state 3: $\mathbf{m}(t) = (0.1816, 0.2500, 0.9511)$.
The leftmost panels show the trajectories of $\mathbf{m}(t)$ for $t \in [0, 80]$.
The middle panels display the time evolution of $|\mathbf{m}(t)|$ and the purity $\mathbf{Tr}[\hat{\rho}^2(t)]$, where the black dashed line denotes the purity of a maximally mixed state, given by $1/L$, which sets the lower bound of $\mathbf{Tr}[\hat{\rho}^2(t)]$.
The rightmost panels show snapshots of the density matrix at $t = 80$ for system size $L = 30$.}
\end{figure*}

In Fig.~\ref{finiteN_dynamics}, we show the dynamics of $m^z(t)$ for both quantum and classical cases for some representative parameters. 
In all panels, we set the interaction strength $(J_x, J_y, J_z)=(0, 1, 0)$ and the initial state to be $x$-polarized. 
Figure~\ref{finiteN_dynamics}(a) shows the dynamics at dissipation strength $\kappa=3$ and a very weak driving strength $\Gamma=0.078$. Under these parameters, the drive is negligible, and the dynamics is governed almost entirely by dissipation and interactions. 
Since the dissipation is implemented via a jump operator [see Eq.~\eqref{jumpoperator}] that continuously annihilates the system in the $z$-direction, the observable $m^z(t)$ rapidly decays to its minimal value of $m^z(t)=-1$, as shown in the plot. 
This behavior is observed both in quantum and classical cases. Following the decay, $m^z(t)$ exhibits weak but non-zero periodic oscillations around $-1$, as shown in the zoomed-in plot, with the same periodicity as the driving. 
The MLE of the thermodynamic-limit system orbit is calculated to be bounded above by zero (as we increase the evolution time, the value gradually approaches zero from negative values), indicating a periodic orbit.

In Fig.~\ref{finiteN_dynamics}(b), we show the dynamics with dissipation strength $\kappa=5$ and driving strength $\Gamma=3.306$. 
Under this choice of parameters, the periodic driving and dissipation compete on a comparable footing, resulting in dynamics that exhibits period-$1$ oscillations in the thermodynamic limit. 
The quantum dynamics show a similar behavior. 
As shown in the zoomed-in plot, the quantum orbits gradually converge toward that of the classical orbit as the system size $N$ increases. The MLE of the classical orbit is calculated to be bounded above by zero, again, indicating a periodic and stable dynamics.

Figure~\ref{finiteN_dynamics}(c) shows the dynamics with $\kappa=0.02$ and $\Gamma=3.306$. 
This parameter point was previously shown in Fig.~\ref{example_finiteN}(b), where the classical dynamics exhibit signatures of irregular, seemingly chaotic behavior. 
We identify this as genuine chaos, as the computed MLE for the thermodynamic limit is $\lambda_{\mathrm{max}}=0.4827>0$. The quantum evolutions converge toward that in the thermodynamic limit only within a short time interval bounded by the Lyapunov time $t=\lambda_{\mathrm{max}}^{-1}\approx 2.07$, as shown in the inset. 
Beyond this short time interval, when the effect of the MLE becomes significant, the quantum and classical orbits diverge substantially, displaying clear quantitative differences. 
We also observe that the quantum evolution of $m^z(t)$ gradually approaches zero and begins to oscillate around this value. 
Under the chosen parameters, the dissipation is very weak, and the dynamics is mainly governed by the all-to-all interactions along the $y$-direction and the periodic driving term. 
Because the finite-size Floquet-Lindblad dynamics admits no strictly symmetry-broken long-time attractor, $m^z(t)$ relaxes toward zero under the combined effect of interactions and dissipation. The periodic drive then sustains a time-periodic response, so that $m^z(t)$ does not remain exactly zero but instead oscillates about zero with the same periodicity.

Figure~\ref{finiteN_dynamics}(d) shows the dynamics with $\kappa=1$ and $\Gamma=8.427$. The dynamical behavior at this parameter point exhibits chaotic dynamics as $\lambda_{\rm max} = 0.0284>0$. 
The quantum dynamics converge toward that of the classical orbit only within the Lyapunov time as $N$ increases. 
Notably, during the time window $10<t<20$, both the quantum and classical orbits appear to attempt an escape from fully chaotic behavior, striving to establish a pattern of periodicity, and the period they attempt to establish is period-$1$, matching the period of the external drive. However, for $t > 20$, although the quantum system continues to form a period-$1$ response, the classical system appears to shift toward period-$2$ dynamics, effectively doubling the period. 
It is important to note that ``periodicity'' here is not exact in the strict sense---like in Fig.~\ref{finiteN_dynamics}(a) and Fig.~\ref{finiteN_dynamics}(b)---but rather reflects an attempt to settle into a regular pattern. After each cycle, the classical system does not return exactly to its previous state, but rather deviates slightly. This behavior is better described as a compromise between chaos and order, which we refer to as quasiperiodic. We will come back to this quasiperiodic behavior later in Sec.~\ref{sec_bifurctaion}.

Figures~\ref{finiteN_dynamics}(e) and~\ref{finiteN_dynamics}(f) show the system's dynamics with $\kappa=3$ and $\Gamma=8.427$, corresponding to the early-time and late-time regimes, respectively. The calculated MLE is $\lambda_{\mathrm{max}}=-0.0009$, indicating a stable behavior for the classical case. We observe that, for the classical system, after discarding the initial transient, the dynamics settles into a stable period-$2$ orbit that persists indefinitely—consistent with the negative MLE. In contrast, the quantum system only briefly attempts to establish a period-$2$ response during the early stage of the evolution $(t<6)$, after which this behavior rapidly decays into a trivial period-$1$ response synchronized with the driving term. This discrepancy in the post-transient periodic behavior arises from the finite size of the quantum system. 
Perturbations such as tunneling in finite systems quickly suppress the period-$2$ dynamics that initially emerge, causing the system to rapidly revert to a trivial period-$1$ response.

We notice that for the stable phase with a negative MLE, shown in Figs.~\ref{finiteN_dynamics}(a) and~\ref{finiteN_dynamics}(b), the quantum and classical orbits exhibit not only qualitatively consistent trends but also quantitative convergence: as the system size increases, the quantum orbit progressively approaches the classical one. In the period-2 dynamics illustrated in Figs.~\ref{finiteN_dynamics}(e) and Figs.~\ref{finiteN_dynamics}(f), although there exist quantitative discrepancies between the quantum and classical evolutions, their qualitative periodic behaviors remain in good agreement—the only difference being that the classical period-2 oscillation is suppressed into a single period in the quantum orbit.
In contrast, as shown in Figs.~\ref{finiteN_dynamics}(c) and~\ref{finiteN_dynamics}(d), in the chaotic and quasiperiodic phases with positive MLEs, the quantum and classical orbits exhibit pronounced deviations. 
The classical orbits display strong chaotic dynamics and quasiperiodic evolution, respectively (as discussed in Sec.~\ref{sec_bifurctaion}, the quasiperiodic phase actually contains intrabranch chaos), whereas the quantum evolution suppresses these nontrivial dynamical behaviors, resulting respectively in a decay to zero and the emergence of periodic motion.

For the chaotic case in Fig.~\ref{finiteN_dynamics}(c), we computed the quantum evolution for different initial states and found that the dynamics of the polarization operator is not confined to the surface of the sphere. 
Instead, the trajectories gradually collapse toward the center of the sphere as the evolution proceeds. This occurs because the quantum system also tends to establish a chaotic behavior, during which the state spreads throughout the entire Hilbert space from a single initial condition, leading to a complete loss of polarization. As illustrated in Fig.~\ref{finiteN_onSphere}, we simulated six different initial states in the quantum system with system size $N=30$ and visualized their trajectories on the sphere. Although the different components of a quantum macrospin do not commute and thus cannot be measured exactly at the same time, the expectation values can still be computed using the density matrix. In the limit as $N\to\infty$ these observables are expected to be able to compare with their classical counterparts. Regardless of the initial condition, all trajectories converge toward the center, as also reflected in the middle panels of Fig.~\ref{finiteN_onSphere}, where the norm of the polarization operator, $|\mathbf{m}(t)|$, rapidly decays from unity to zero. 

Simultaneously, we observe that the \textbf{system’s purity}, $\mathbf{Tr}[\hat{\rho}^2(t)]$, exhibits a synchronized decay—from the pure-state value of 1 down to a small value corresponding to a mixed state, bounded below by the purity of a maximally mixed state, $1/L$. The fact that the purity approaches this lower bound indicates that the system has effectively expanded over the entire Hilbert space, representing a manifestation of quantum chaos. This extended nature is also visible from the density matrix itself: as shown in the rightmost panels of Fig.~\ref{finiteN_onSphere}, where we directly visualize the matrix elements of the density matrices at $t=80$ (we introduce how to visualize the density matrix in Sec.~\ref{Sec_densitymatrix}), the nonzero elements are mainly concentrated along the diagonal. This indicates that the density matrix is approximately a linear combination of the outer products of nearly all basis states, which further demonstrates the extended character of the system in this regime.

Our results reveal that, beyond the transient regime, the quantum and classical dynamics exhibits fundamentally different behaviors in the chaotic region. The deviation between the two is not merely a consequence of the positive MLE and classical chaos, but rather a manifestation of quantum chaos itself. In the quantum system, chaotic evolution leads to a delocalization of the state over the entire Hilbert space, resulting in a loss of polarization and the decay of $\abs{\mathbf{m}(t)}$ toward zero. This global spreading is therefore not a suppression of chaos, but its quantum counterpart---reflecting how chaotic information becomes fully distributed in the state space.

\section{Density matrix profile}\label{Sec_densitymatrix}

\begin{figure*}[ht]
\centering
\includegraphics[width=0.97\textwidth]{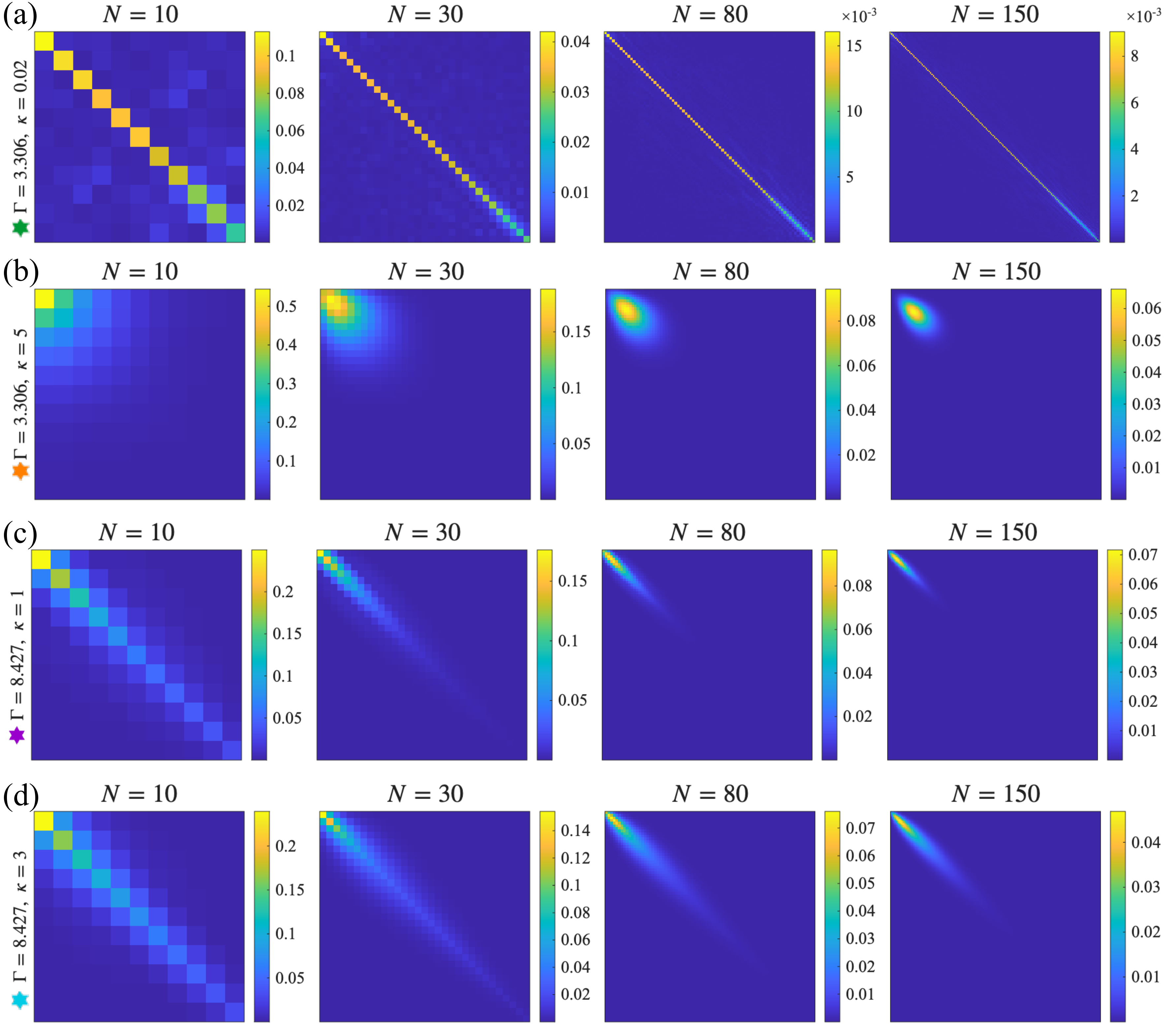}
\caption{The density-matrix element profiles of the quantum case in (a) chaotic, (b) 1-periodic, (c) quasiperiodic, and (d) 2-periodic regimes, respectively. For a given system size $N$, the density matrix has dimension $(N+1)\times(N+1)$. Each panel visualizes the density matrix, with the color scale indicating the absolute value of each matrix element.}
\label{densitymatrixelement}
\end{figure*}

In Ref.~\cite{PhysRevA.109.013328}, the authors argued that the disagreement observed between classical and quantum evolutions at certain parameter values originates from a breakdown of the classical (mean-field) equations in that regime, so that these equations no longer correctly describe the true thermodynamic-limit dynamics. 
According to their analysis, this putative breakdown can be diagnosed via the localization properties of the quantum density matrix in the Dicke basis, $\rho_{M,M'} = \mel{M}{\rho}{M'}$, viewed in the $\frac{2M}{N}$–$\frac{2M'}{N}$ plane. 
They proposed that the classical description remains valid only when the nonzero matrix elements of $\hat{\rho}$ are sharply localized within a narrow strip of width $\sim N^{-1/2}$.
However, according to the analytical results presented in Ref.~\cite{PhysRevLett.133.150401}, at any given time, as long as the system size is sufficiently large, the quantum evolution will always converge to the thermodynamic-limit orbit predicted by the classical mean-field approach. In other words, the classical equations consistently provide an accurate description of the system’s dynamics in the thermodynamic limit, and the observed disagreement arises entirely from strong finite-size effects in the quantum case. By considering large enough values of $N$ in the quantum system, one can always obtain orbits that converge toward the classical results.

We argue that, although there is no ``breakdown'' for the classical mean field theory, the ``localization'' behavior of the density matrix can indeed serve as an indicator of the consistency between quantum and classical evolutions. 
When discrepancies arise, the underlying reason is not the breakdown of the classical equations, but rather the failure of the quantum orbits due to finite-size effects in the quantum system. According to the theorem proved in Ref.~\cite{PhysRevLett.133.150401}, the error function $\mathcal{E}$ between quantum and classical evolutions is bounded by a function of system size $N$ and evolution time $t$, given by 
\begin{align}
    \mathcal{E}_N(t) < \frac{C_1}{N}e^{C_2t},
\end{align}
where $C_1$ and $C_2$ are $N$-independent constants. Furthermore, as discussed in Sec.~\ref{sec_finiteNandinfiniteN}, we find that in the chaotic phase, the nonzero elements of the density matrix are concentrated along the diagonal, which in fact signifies the establishment of quantum chaos.

In Fig.~\ref{densitymatrixelement}, we compute the density matrix profiles of the system in different dynamical phases. As discussed in Sec.~\ref{sec_finiteNandinfiniteN}, the system’s dynamical behavior can be regarded as stabilized once the initial transient regime with $t<20$ is neglected. We therefore consider snapshots of the density matrix at $t=80$. The parameter combinations chosen in Figs.~\ref{densitymatrixelement}(a), \ref{densitymatrixelement}(b), \ref{densitymatrixelement}(c), and \ref{densitymatrixelement}(d) correspond to the chaotic, period-1, quasiperiodic, and period-2 dynamical phases, respectively.

In the chaotic phase [Fig.~\ref{densitymatrixelement}(a)], the density matrix exhibits nonzero projection across all basis states, independent of the system size $N$. Consequently, the system explores the entire state space during its evolution, giving rise to highly unpredictable chaotic dynamics. In this regime, the positive MLE renders the classical orbit ill-defined, and therefore the quantum evolution is not expected to converge toward it. At this parameter point, the quantum and classical orbits show strong disagreement [see Fig.~\ref{finiteN_dynamics}(c)], as expected. However, this disagreement does not imply a fundamental difference in the nature of the dynamics. 
Both the quantum and classical systems exhibit chaotic behavior, although it manifests in distinct ways in each case.

In Fig.~\ref{densitymatrixelement}(b), the nonzero elements of the density matrix are concentrated near a small subset of basis states, and this concentration becomes sharper as $N$ increases, demonstrating a clear localization. At this parameter point, both the quantum and classical dynamics exhibit trivial period-1 behavior, and the quantum orbits converge progressively toward the classical one as $N$ grows [see Fig.~\ref{finiteN_dynamics}(b)].

For Figs.~\ref{densitymatrixelement} (c) and \ref{densitymatrixelement}(d), at small $N$, the density matrix has nonzero projections over a wide range of basis states. As $N$ increases, the number of basis states with nonzero projections decreases, but no clear localization behavior emerges in the density matrix. 
These cases correspond to quasiperiodic [Fig.~\ref{densitymatrixelement}(c)] and period-2 [Fig.~\ref{densitymatrixelement}(d)] dynamics, respectively, in which the quantum and classical orbits both display noticeable disagreement [see Figs.~\ref{finiteN_dynamics}(d) and (f)]. 
Although the density matrix profiles appear similar in these two cases, the origins of the discrepancies in the quasiperiodic and period-2 phases may be fundamentally different. 
For the quasiperiodic phase in Fig.~\ref{densitymatrixelement}(c), the system possesses a positive MLE. Despite exhibiting an approximate period-2 structure, the dynamics within each branch are essentially chaotic and thus unpredictable. In this situation, the classical evolution does not possess a well-defined orbit, let alone one to which the quantum orbits could be expected to converge. Therefore, just as in the genuinely chaotic case shown in Fig.~\ref{densitymatrixelement}(a), the discrepancy between the quantum and classical dynamics here arises from the unpredictability induced by the positive MLE. However, the partially delocalized nature of the density matrix at this stage indicates that the system possesses a certain degree of quantum chaos, which is consistent with the chaotic behavior induced by the positive MLE in the classical counterpart.

For the period-$2$ case shown in Fig.~\ref{densitymatrixelement}(d), the negative MLE implies that small perturbations should decay rapidly, so one would expect the quantum orbit to converge toward the classical one. However, at finite size the evolution is inevitably influenced by effects such as tunneling. Consequently, although the quantum dynamics may transiently follow the classical period-$2$ structure, this behavior is quickly destabilized by finite-size fluctuations, and the system crosses over to a trivial period-$1$ response. Since the ensuing delocalization of the density matrix is therefore a finite-size effect, we expect it to be progressively suppressed as the system size increases, leading to the recovery of localization in the large-$N$ limit.

\section{Conclusions}\label{sec_conclusion}
Overall, the reported study of nonequilibrium macrospin dynamics of a periodically driven, dissipative multi-spin systems with all-to-all interactions, demonstrates the effectiveness of combining mean-field theory with Lyapunov exponents analysis (e.g., by using the maximal Lyapunov exponent, MLE, to quantify the stability of the evolution orbit) for systematically classifying their complex behavior on the routes to chaos and making a comparison between the quantum case and the classical cases.  

For the thermodynamic ($N \to \infty$) limit, we provide a detailed characterization of the system’s dynamical phases through bifurcation diagrams, Lyapunov analysis, and Fourier spectra for observables. Our algorithm for computing $\lambda_{\mathrm{max}}$ is proven to be both stable and reliable across a wide range of parameters, and served as a key diagnostic throughout. We observed apparent discontinuities in bifurcation diagrams that were not associated with sharp phase transitions, but instead arose from continuous deformations of the system’s regions of attractors. In certain regimes, these regions exhibit fractal-like boundaries, further emphasizing the complexity of the system’s nonlinear dynamics. In addition, we identified a rich dynamical phase structure, including period-doubling bifurcations culminating in chaos, with scaling behavior consistent with the first Feigenbaum constant. 

By simulating quantum systems using the permutationally invariant basis and comparing their orbits to the mean-field classical dynamics in the thermodynamic limit, we found that the correspondence between classical and quantum dynamics strongly depends on the sign of the MLE, $\lambda_{\mathrm{max}}$. When $\lambda_{\mathrm{max}} < 0$, the classical system lies in a stable phase, and the quantum orbits reliably converge toward the classical result even after long-time evolution. For classical dynamics with nontrivial periodic attractors (e.g., period-2 orbits), the corresponding quantum system initially attempts to follow the same periodic pattern, but, eventually, decays into a very different 'trivial' period-1 dynamics, due to the finite-N effects which cause quantum tunneling of the macrospin. 

Furthermore, when $\lambda_{\mathrm{max}} > 0$, the classical system is chaotic (orbits become ill-defined due to exponential sensitivity to any perturbation), quantum evolution cannot be linked to any certain orbit. That apparent mismatch between classical and quantum dynamics does not imply any fundamental difference in the underlying chaotic character of the quantum and classical systems. In fact, both exhibit unambiguous signatures of chaos. In the classical system, the chaotic behavior is evident from the highly unpredictable orbits and the positive MLE. In the quantum counterpart, however, an examination of the density matrix reveals that its nonzero matrix elements are predominantly concentrated along the diagonal, signifying that the system evolves from a localized initial state to a diffusive spreading over the entire Hilbert space, thus, capturing the essence of quantum chaos. 
This delocalization leads to the loss of polarisation, which in turn explains the absence of agreement with the classical orbits.

\begin{acknowledgments}
XL is supported by the Research Grants Council of Hong Kong (Grants No. CityU 11300421, CityU 11304823, C7012-21G, and C7015-24G) and City University of Hong Kong (Project No. 9610428). VF and HF thank the Distinguished Visitor program of CityUHK for arranging international mobility and cooperation between CityUHK and the University of Manchester. 
\end{acknowledgments}

\appendix
\section{Further discussion on the single-particle model}\label{secA1}
The single-particle Hamiltonian is given by
\begin{align}\label{singleparticleHamiltonian}
    H_{\mathrm{sp}}(t) &=\frac{\Gamma}{2}\sin{(\omega t)} \hat{\sigma}^x + \frac{\Gamma}{2}[1-\cos{(\omega t)}] \hat{\sigma}^y\nonumber\\
    &= \Gamma\sin({\frac{\omega t}{2}})[\cos(\frac{\omega t}{2})\hat{\sigma}^x + \sin(\frac{\omega t}{2})\hat{\sigma}^y]
\end{align}
which has been studied for both the noninteracting case~\cite{fu2022quantum,PhysRevResearch.5.L032024} and the case with nearest-neighbor Ising interaction~\cite{PhysRevResearch.5.L032024}. It is known to possess a nonsymmorphic glide symmetry $\widetilde g$, with its matrix representation denoted by $G(t)$, such that $[H_{\mathrm{sp}}(t), G(t)]=0$. Analytically, the two instantaneous eigenstates, the corresponding eigenvalues of the Hamiltonian~\ref{singleparticleHamiltonian} and the matrix representation $G(t)$ are given by
\begin{align}\label{SP_eigenstates}
    \ket{\pm, t} &= \frac{1}{\sqrt{2}}[\pm e^{-i\omega t},1]^{\mathsf{T}},\nonumber\\
    E_{\pm}(t) &= \pm\Gamma\sin{(\frac{\omega t}{2})},\nonumber\\
    G(t) &= {\begin{pmatrix} 0 & e^{-i\omega t} \\ 1 & 0 \end{pmatrix}}.
\end{align}
Notice that the two states are gapless at integer multiples of period $T$. In fact, it is the symmetry $\widetilde g$ that forces the two instantaneous eigenstates of the single-particle Hamiltonian~\ref{singleparticleHamiltonian} to be gapless. In addition, the period of instantaneous eigenstates is doubled with respect to that of the single-particle Hamiltonian~\ref{singleparticleHamiltonian} due to the M\"obius twist of the two instantaneous energy eigenstates.

\subsection{$\widetilde g$-induced gapless band}
In fact, Hamiltonians that commute with the representation matrix $G(t)$ of the nonsymmorphic symmetry $\widetilde g$ is guaranteed to be gapless.

Assume a general form of a periodic Hamiltonian
\begin{align}
     H(t) &= {\begin{pmatrix} a(t) & b(t) \\ b^*(t) & c(t)\end{pmatrix}}
\end{align}
with period $T=2\pi/\omega$, that commutes with 
\begin{align}\label{G_matrix}
     G(t) &= {\begin{pmatrix} 0 & e^{-i\omega t} \\ 1 & 0 \end{pmatrix}}.
\end{align}
We have
\begin{align}
    [H(t), G(t)] = {\begin{pmatrix} b(t)-b^*(t)e^{-i\omega t} & (a-c)e^{-i\omega t} \\ c-a &  b^*(t)e^{-i\omega t}-b(t)\end{pmatrix}} = 0,
\end{align}
which indicates $a=c$ and
\begin{align}\label{bt_relation}
    b(t) = b^*(t)e^{-i\omega t},
\end{align}
and the Hamiltonian can be written by
\begin{align}
    H(t) = {\begin{pmatrix} 0 & b(t) \\ b^*(t) &  0\end{pmatrix}}
\end{align}
via a global shift of the spectrum. 

To prove the two instantaneous eigenstates of the Hamiltonian are gapless somewhere, it suffices to show that $b(t)$ vanishes at some time.  We begin by defining a new function $z(t) = b(t)e^{i\omega t/2}$.. By multiplying both sides of Eq.~\eqref{bt_relation} by $e^{i\omega t/2}$, it follows that $z(t)$ is real-valued. Therefore, we can express $b(t)$ as $b(t) = z(t)e^{-i\omega t}$. Since $b(t)$ is periodic with period $T=2\pi/\omega$, we have $b(t+T)=b(t)$ which implies $z(t+T)=-z(t)$. That is, $z(t)$ is anti-periodic. Because $z(t)$ is a continuous real-valued function and is anti-periodic, the intermediate value theorem guarantees that $z(t)=0$ for some $t$. Consequently, $b(t)$ also vanishes at that point. This implies that the Hamiltonian exhibits a level crossing (i.e., becomes gapless) at some time.

\subsection{The family of $\widetilde g$-compatible Hamiltonians}
We further assume that $b(t)$ is purely imaginary and is an even function for $t$. We have the discrete Fourier transform $b(t)=\sum_n b_n e^{in\omega t}$, in which one can easily proof $b^*_n=-b_n$ from the assumption. 

Via complex conjugate we have $b^*(t)=\sum_n b^*_n e^{-in\omega t}$, and from Eq.~\eqref{bt_relation} we have $b^*(t) = b(t)e^{i\omega t}=\sum_n b_n e^{i(n+1)\omega t}$. Comparing the two equations, we have $b_{-(n+1)} = b^*_n$. Next, we split the infinite series into two parts by
\begin{align}
    b(t)&=\sum_{n=0}^{+\infty}b_ne^{in\omega t} + \sum_{n=-1}^{-\infty}b_ne^{in\omega t}\nonumber\\
    &=\sum_{n=0}^{+\infty}b_ne^{in\omega t} + \sum_{n=0}^{+\infty}b^*_ne^{-i(n+1)\omega t}\nonumber\\
    &=\sum_{n=0}^{+\infty}b_n\big[e^{in\omega t }-e^{-i(n+1)\omega t} \big]\nonumber\\
    &=\sum_{n=0}^{+\infty}\frac{r_n}{i}\big[e^{in\omega t }-e^{-i(n+1)\omega t} \big]
\end{align}
where $r_n$ are real numbers. With this general form of $b(t)$, we have a family of Hamiltonians given by
    \begin{align}\label{H_family}
    H(t)=& \notag\\
    \sum_{n=0}^{+\infty} 
    &{\begin{pmatrix} 0 & \frac{r_n}{i}\big[e^{in\omega t }-e^{-i(n+1)\omega t} \big] \\ 
    \frac{r_n}{-i}\big[e^{-in\omega t }-e^{i(n+1)\omega t} \big] & 0\end{pmatrix}}.
    \end{align}
We get different Hamiltonians by setting different values of $r_n$ for all $n$. In particular, if we let all $r_{n\neq0}=0$, and $r_0=\frac{\Gamma}{2}$, we get the single-particle Hamiltonian~\ref{singleparticleHamiltonian} we are considering. In addition, the general form of Hamiltonian~\ref{H_family} also commutes with the nonsymmorphic symmetry~\ref{G_matrix}, and takes the two time-dependent single-particle states given in Eq.~\eqref{SP_eigenstates} as its instantaneous eigenstates as well. 

In this paper, we focus on the simplest Hamiltonian in this family, which already exhibits remarkably rich dynamical behavior. It is foreseeable that choosing other Hamiltonians with more nonzero $r_n$ terms would lead to even more complex dynamical phenomena.

\section{Derivation for the finite-size quantum system}\label{secA2}
We introduce how to derive the Lindblad Eq.~\eqref{finiteNLindblad} and the expression for the initial state (\ref{finiteNInitialstate}) in the main text.
\subsection{Finite-$N$ Lindblad Eq.~\eqref{finiteNLindblad}}
We start from the creation and annihilation relations in~\ref{crea_anni} shown in the main text, which are obtained by directly left multiplying the collective spin operators on the basis state given in~\ref{Dickebasis}. Specifically, we have
\begin{widetext}
\begin{align}
    \hat{S}^z\ket{M} &= \frac{2M}{N}\ket{M},\\
    \hat{S}^x\ket{M} &= \frac{1}{N} \sqrt{\frac{1}{{C_{N}^{N/2+M}}}} \Bigg[\bigg(\frac{N}{2}-M+1\bigg)\bigg( \sum_{\sum_j s_j^z = M-1} \ket{s_1^z, s_2^z, \cdots, s_N^z} \bigg) + \bigg(\frac{N}{2}+M+1\bigg)\bigg(\sum_{\sum_j s_j^z = M+1} \ket{s_1^z, s_2^z, \cdots, s_N^z}\bigg)\Bigg]\nonumber\\
    &=\frac{1}{N}\bigg(\frac{N}{2}-M+1\bigg)\sqrt{\frac{{C_{N}^{N/2+M-1}}}{{C_{N}^{N/2+M}}}}\ket{M-1} + \frac{1}{N}\bigg(\frac{N}{2}+M+1\bigg)\sqrt{\frac{{C_{N}^{N/2+M+1}}}{{C_{N}^{N/2+M}}}}\ket{M+1}\nonumber\\
    &=\frac{1}{N}\sqrt{(\frac{N}{2}-M+1)(\frac{N}{2}+M)}\ket{M-1} +\frac{1}{N}\sqrt{(\frac{N}{2}+M+1)(\frac{N}{2}-M)}\ket{M+1} \nonumber\\
    &=\mathcal{F}_M\ket{M-1} + \mathcal{F}_{M+1}\ket{M+1},\nonumber\\
    \hat{S}^y\ket{M} &= \frac{1}{N} \sqrt{\frac{1}{{C_{N}^{N/2+M}}}} \Bigg[i\dfrac{N - 2M + 2}{2}\sum_{\sum_{j=1}^N \sigma_j^z = M-1} \ket{\sigma_1^z, \sigma_2^z, \cdots, \sigma_N^z} - i\dfrac{N + 2M + 2}{2}\sum_{\sum_{j=1}^N \sigma_j^z = M+1} \ket{\sigma_1^z, \sigma_2^z, \cdots, \sigma_N^z}\Bigg]\nonumber\\
    &=\frac{i}{N}\bigg(\frac{N}{2}-M+1\bigg)\sqrt{\frac{{C_{N}^{N/2+M-1}}}{{C_{N}^{N/2+M}}}}\ket{M-1} - \frac{i}{N}\bigg(\frac{N}{2}+M+1\bigg)\sqrt{\frac{{C_{N}^{N/2+M+1}}}{{C_{N}^{N/2+M}}}}\ket{M+1}\nonumber\\
    &=\frac{i}{N}\sqrt{(\frac{N}{2}-M+1)(\frac{N}{2}+M)}\ket{M-1} -\frac{i}{N}\sqrt{(\frac{N}{2}+M+1)(\frac{N}{2}-M)}\ket{M+1} \nonumber\\
    &=i\mathcal{F}_M\ket{M-1} - i\mathcal{F}_{M+1}\ket{M+1}\notag, 
\end{align}
where $\mathcal{F}_M = \frac{1}{N}\sqrt{(\frac{N}{2}-M+1)(\frac{N}{2}+M)} = \sqrt{(\frac{1}{2}+\frac{M}{N})(\frac{1}{2}-\frac{M}{N}+\frac{1}{N})}$ is a function of $M$.

Then, we can multiply the collective spin operators to the density matrix. We have
\begin{align}
    S^x \sum_{M, M'}\rho_{M, M'}\ket{M}\bra{M'} &= \sum_{M, M'}\rho_{M, M'}\bigg(\mathcal{F}_M\ket{M-1}+\mathcal{F}_{M+1}\ket{M+1}\bigg)\bra{M'}\nonumber\\
    &=\sum_{M, M'}\mathcal{F}_{M}\rho_{M, M'}\ket{M-1}\bra{M'} + \sum_{M, M'}\mathcal{F}_{M+1}\rho_{M, M'}\ket{M+1}\bra{M'}\nonumber\\
    &=\sum_{M, M'}\mathcal{F}_{M+1}\rho_{M+1, M'}\ket{M}\bra{M'}+\sum_{M, M'}\mathcal{F}_M\rho_{M-1, M'}\ket{M}\bra{M'}\nonumber\\
    &=\sum_{M, M'}\bigg(\mathcal{F}_{M+1}\rho_{M+1, M'}+\mathcal{F}_M\rho_{M-1, M'}\bigg)\ket{M}\bra{M'},
\end{align}
\end{widetext}
and similar results can be derived for $S^y$ and $S^z$. We insert all these equations into the Lindblad Eq.~\eqref{LindbladEq}, and thus we obtain the Lindblad Eq.~\eqref{finiteNLindblad} for finite-size quantum system.

\subsection{Finite-$N$ initial state}
We consider the initial state as a  product state with all the spins aligned in the same direction represented by the azimuthal angles $(\theta, \phi)$, with $\theta\in[0, \pi)$, $\phi\in[0,2\pi)$. For one spin aligned with $z$-direction given by $\ket{\uparrow}=\ket{(\theta=0, \phi=0)}$, it can be rotated to the angle $(\theta, \phi)$ by the rotational matrix $\mathcal{R}(\theta, \phi)$, and precisely, given by
\begin{align}
   \mathcal{R}(\theta, \phi)\ket{\uparrow} = \cos{\frac{\theta}{2}}\ket{\uparrow}+e^{i\phi}\sin{\frac{\theta}{2}}\ket{\downarrow}.
\end{align}
Thus the spin chain with all spins aligned to the angle $(\theta, \phi)$ is given by
\begin{align}
    &\ket{\theta, \phi}=\underset{j=1}{\overset{N}{\bigotimes}}\bigg(\cos{\frac{\theta}{2}}\ket{\uparrow}_j+e^{i\phi}\sin{\frac{\theta}{2}}\ket{\downarrow}_j\bigg)\nonumber\\
    &=\sum_{k=0}^{N}(\cos{\frac{\theta}{2}})^k (e^{i\phi}\cos{\frac{\theta}{2}})^{N-k}\ket{k\uparrow, (N-k)\downarrow}\nonumber\\
    &=\sum_{M=-\frac{N}{2}}^{\frac{N}{2}}(\cos{\frac{\theta}{2}})^{\frac{N}{2}+M} (e^{i\phi}\sin{\frac{\theta}{2}})^{\frac{N}{2}-M}\sqrt{C_N^{N/2+M}}\ket{M},
\end{align}
where we notice that $M=k\times\frac{1}{2} + (N-k)\times(-\frac{1}{2})=k-\frac{N}{2}$.

Hence, we can initialize the finite-size system with all spins aligned to the angle $(\theta, \phi)$ by setting the density matrix elements by
\begin{align}
    &\rho_{M, M'}(0)=\bigg(\ket{\theta, \phi}\bra{\theta, \phi}\bigg)_{M, M'}\nonumber\\
    =&\sqrt{C_N^{N/2+M}C_N^{N/2+M'}}(\cos{\frac{\theta}{2}})^{N+M+M'}(\sin{\frac{\theta}{2}})^{N-M-M'}\nonumber\\
    &\times e^{i\phi(M'-M)}\ket{M}\bra{M'}.
\end{align}

\subsection{Symmetry-induced dynamical constraint}
\begin{figure}[b]
\centering
\includegraphics[width=0.48\textwidth]{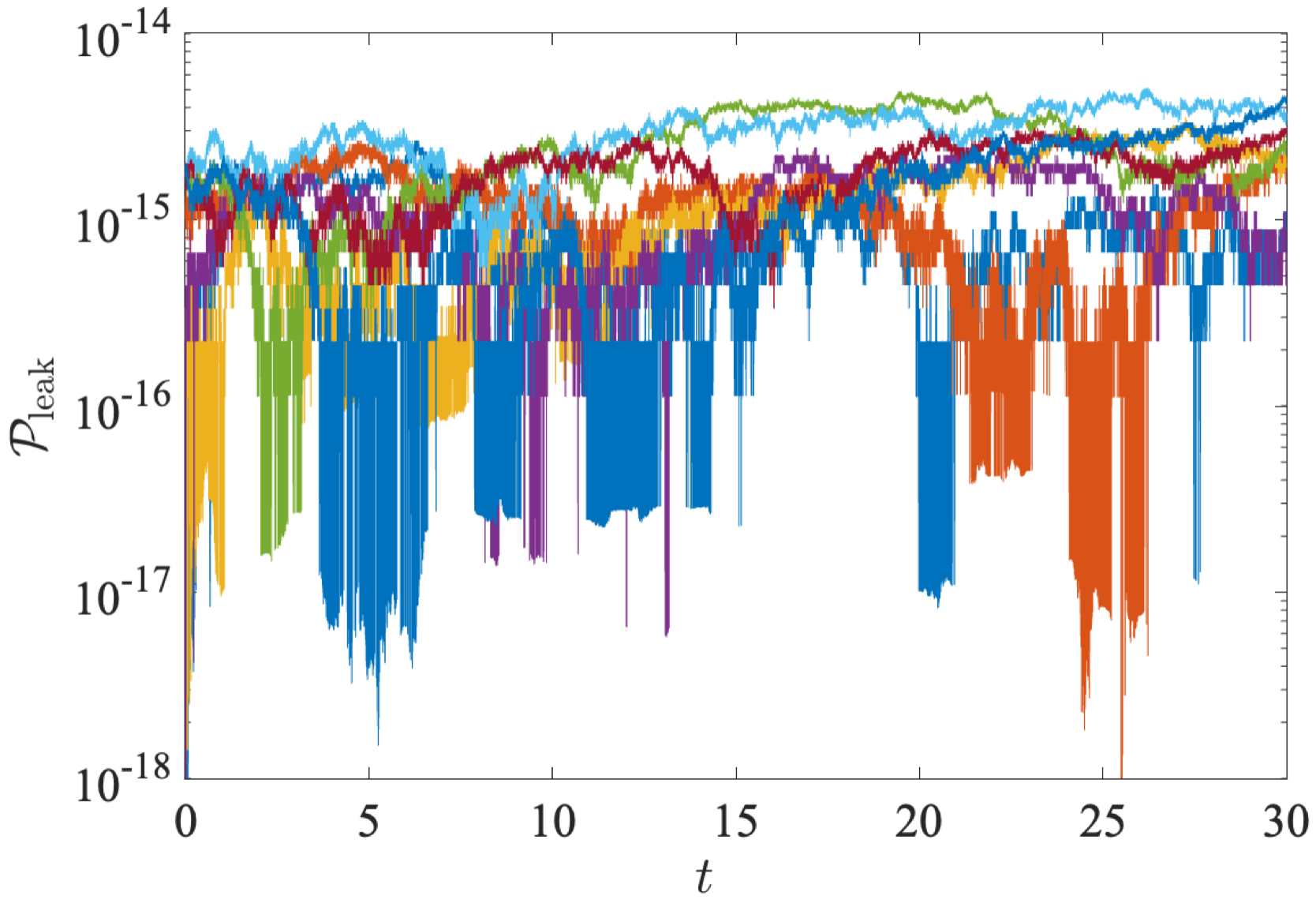}
\caption{Evolution of the leak for eight randomly chosen initial states. The parameters: $\Gamma$, $\kappa$, and $J_{\mu}$ $(\mu=x,y,z)$ are also randomly generated from the parameter space.\label{finiteNleak}}
\end{figure}

Since the Hamiltonian~\ref{manybodyHamiltonian} and the jump operator~\ref{jumpoperator} have the permutational invariance, the density matrix $\hat{\rho}(t)$ is always restricted in the permutationally-invariant Dicke subspace. This can be verified by measuring the projection of the density matrix onto the orthogonal complement of this space. The projection operator onto the permutationally-invariant subspace is $P_{\mathrm{Dicke}}=\sum_{M=-\frac{N}{2}}^{\frac{N}{2}}\ket{M}\bra{M}$. Then, the orthogonal complement operator which measures the leak is given by $P_{\mathrm{leak}}=I-P_{\mathrm{Dicke}}$, and for any density matrix $\hat{\rho}(t)$, its leak probability is
\begin{align}
    \mathcal{P}_{\mathrm{leak}} = {\mathbf{Tr}}(P_{\mathrm{leak}}\hat{\rho}(t)) = 1-{\mathbf{Tr}}(P_{\mathrm{Dicke}}\hat{\rho}(t)).
\end{align}
We find that the leak remains zero up to machine precision limit ($\sim10^{-15}$) for any choice of parameters and initial states, as illustrated in Fig.~\ref{finiteNleak}. The eight curves shown in the figure are merely representative examples, but they capture general features of the system’s behavior across the whole range of parameters.

\section{Symmetry analysis for quantum system}

We provide an analysis of the permutational symmetry together with the $SU(2)$ spin rotational symmetry~\cite{fulton1997young, edmonds1996angular}. In this section, a slightly different notation is used compared to other sections for the convenience of mathematical exposition. To avoid overly abstract mathematics, we will use a system composed of six spin-$\frac{1}{2}$ particles as an example for the subsequent discussion.

\subsection{Hilbert space and the permutation representation}

For each spin-$\tfrac{1}{2}$ particle, the local Hilbert space is
\begin{align}
    \mathbb{C}^2 = \mathrm{span}\{\ket{\uparrow}, \ket{\downarrow}\}.
\end{align}
For six spins, the total Hilbert space is
\begin{align}
    \mathcal{H} = (\mathbb{C}^2)^{\otimes 6},
\end{align}
with dimension
\begin{align}
    \dim \mathcal{H} = 2^6 = 64.
\end{align}

For $N$ spin-$\tfrac{1}{2}$ particles, the possible total spins are
\begin{align}
J = \frac{N}{2},\ \frac{N}{2}-1,\ \dots,\ J_{\min},
\end{align}
with $J_{\min}=0$ when $N$ is even and $J_{\min}=\frac{1}{2}$ when $N$ is odd.

For $N=6$ (even), we have
\begin{align}
    J \in \{3,2,1,0\}.
\end{align}

An orthonormal basis is given by the states
\begin{align}
    \ket{s_1 s_2 s_3 s_4 s_5 s_6}
= \ket{s_1} \otimes \ket{s_2}\otimes \ket{s_3} 
  \otimes \ket{s_4} \otimes \ket{s_5} \otimes \ket{s_6},
\end{align}
where $s_i \in \{\uparrow,\downarrow\}$.

\begin{definition}
For each permutation $\sigma \in S_6$, define $P_\sigma$ on $\mathcal{H}$ by
\begin{align}
P_\sigma\bigl(\ket{s_1}\otimes\cdots\otimes\ket{s_6}\bigr)=\ket{s_{\sigma^{-1}(1)}}\otimes\cdots\otimes\ket{s_{\sigma^{-1}(6)}}.
\end{align}
\end{definition}
One checks that $P_{\sigma_1}P_{\sigma_2} = P_{\sigma_1\sigma_2}$, so $\sigma \mapsto P_\sigma$ is a representation of $S_6$ on $\mathcal{H}$. Physically, $P_\sigma$ simply permutes the labels of the spins.

We are interested in Hamiltonians $H$ satisfying
\begin{equation}
[H,P_\sigma] = 0,\  \forall \sigma \in S_6,
\label{eq:perm-sym-H}
\end{equation}
i.e. $H$ is invariant under all particle permutations.
Such Hamiltonians arise, for example, in fully connected spin models, where interaction terms are symmetric under exchanging any pair of spins, just like the model we are considering.

\subsection{Schur--Weyl duality}

\subsubsection{$SU(2)$ and $S_6$ acting on $(\mathbb{C}^2)^{\otimes 6}$}

Two group actions commute on $\mathcal{H}$:

\begin{itemize}
  \item $SU(2)$ acts via the tensor product of the representation:
  \begin{align}
      U^{\otimes 6}, \  U \in SU(2).
  \end{align}
  \item $S_6$ acts via $P_\sigma$ as defined above.
\end{itemize}

These actions commute:
\begin{align}
    U^{\otimes 6} P_\sigma = P_\sigma U^{\otimes 6},\  \forall U\in SU(2),\ \sigma\in S_6.
\end{align}

\subsubsection{Schur--Weyl duality}

Schur--Weyl duality states that the joint action of $SU(2)$ and $S_6$ on $\mathcal{H}$ yields a decomposition
\begin{equation}
\mathcal{H} \cong \bigoplus_{J} \left(V_J \otimes W_J\right),
\label{eq:SW-decomp}
\end{equation}
where:

\begin{itemize}
  \item $V_J$ is the irreducible representation (irrep) of $SU(2)$ of total spin $J$, with dimension
  \begin{align}
  \dim V_J = 2J+1.
  \end{align}
  \item $W_J$ is an irrep of $S_6$, labeled by a Young diagram $\lambda(J)$, with dimension
  \begin{align}
  \dim W_J = f^{\lambda(J)},
  \end{align}
  which is also the multiplicity of $V_J$ in $\mathcal{H}$.
\end{itemize}

\subsection{Young diagrams for $N=6$}

\subsubsection{Young diagrams representations}

For $N$ spin-$\tfrac{1}{2}$ particles, the $S_N$ irreps that appear have Young diagrams with at most two rows, and the diagram associated with total spin $J$ is
\begin{align}
    \lambda(J) = \left(\frac{N}{2}+J,\ \frac{N}{2}-J\right).
\end{align}

For $N=6$, we have $\frac{N}{2}=3$, so
\begin{align}
    \lambda(J) = (3+J,\ 3-J).
\end{align}
Thus:
\begin{align*}
    J=3 &\Rightarrow \lambda(3) = (6,0) \equiv (6),\\
    J=2 &\Rightarrow \lambda(2) = (5,1),\\
    J=1 &\Rightarrow \lambda(1) = (4,2),\\
    J=0 &\Rightarrow \lambda(0) = (3,3).
\end{align*}
The corresponding Young diagrams are given by:

\begin{itemize}
  \item $(6)$ (one row of 6 boxes, fully symmetric representation):
  \begin{align}
  \YoungSix
  \end{align}
  \item $(5,1)$ (first row 5 boxes, second row 1 box):
  \begin{align}
  \YoungFiveOne
  \end{align}
  \item $(4,2)$ (first row 4 boxes, second row 2 boxes):
  \begin{align}
  \YoungFourTwo
  \end{align}
  \item $(3,3)$ (first row 3 boxes, second row 3 boxes):
  \begin{align}
  \YoungThreeThree
  \end{align}
\end{itemize}

\subsection{Dimensions of the $S_6$ irreps}

The dimension $f^\lambda$ of the irreducible representation labeled by Young diagram $\lambda$ can be computed via the hook-length formula. For the four shapes above, one finds:
\begin{align}
f^{(6)} = 1,\ 
f^{(5,1)} = 5,\ 
f^{(4,2)} = 9,\ 
f^{(3,3)} = 5.
\end{align}

\subsection{Full decomposition for $N=6$}

Combining Schur--Weyl duality with these dimensions, we have
\begin{equation}
(\tfrac{1}{2})^{\otimes 6}
\cong
V_3 \otimes W_{(6)}
\ \oplus\
V_2 \otimes W_{(5,1)}
\ \oplus\
V_1 \otimes W_{(4,2)}
\ \oplus\
V_0 \otimes W_{(3,3)}.
\label{eq:full-decomp}
\end{equation}
The dimensions are summarized in Table~\ref{tab:decomp}.

\begin{table*}[ht]
\centering
\begin{tabular}{c|c|c|c|c}
\hline\hline
$J$ & $\dim V_J = 2J+1$ & Young diagram $\lambda(J)$ & $\dim W_{\lambda(J)} = f^{\lambda(J)}$ & total dimension \\
\hline
3 & 7 & $(6)$     & $1$ & $7$ \\
2 & 5 & $(5,1)$   & $5$ & $25$ \\
1 & 3 & $(4,2)$   & $9$ & $27$ \\
0 & 1 & $(3,3)$   & $5$ & $5$ \\
\hline\hline
\end{tabular}
\caption{Decomposition of the six-spin Hilbert space into $SU(2)\times S_6$ sectors.}
\label{tab:decomp}
\end{table*}

Summing the total dimensions:
\begin{align}
7 + 25 + 27 + 5 = 64 = 2^6,
\end{align}
as expected.

\subsection{The fully symmetric subspace and Dicke states}

\subsubsection{The fully symmetric sector: $J=3$, $\lambda=(6)$}

The Young diagram $(6)$ represents the completely symmetric representation of $S_6$. Its dimension is $f^{(6)} = 1$, meaning there is a unique (up to equivalence) such symmetry type in $\mathcal{H}$.

This symmetric sector corresponds to the maximal angular momentum sector with total spin
\begin{align}
    J = 3,
\end{align}
with $SU(2)$ irrep dimension
\begin{align}
    \dim V_3 = 2\cdot 3 + 1 = 7.
\end{align}
Thus the fully symmetric subspace has dimension
\begin{align}
    \dim \mathcal{H}_\text{sym} = 7.
\end{align}

Physically, this is the ``collective spin $J=3$'' subspace of six spin-$\tfrac{1}{2}$ particles, and it is natural to describe it using Dicke states.

\subsubsection{Dicke states for six spins: general definition}

For $N=6$, the Dicke state with exactly $k$ spins up and $6-k$ spins down is defined as
\begin{equation}
\ket{D_k} = \frac{1}{\sqrt{\binom{6}{k}}}
\sum_{\substack{s_1,\dots,s_6 \\ \#\{\uparrow\}=k}}
\ket{s_1 s_2 s_3 s_4 s_5 s_6},
\  k=0,1,\dots,6.
\label{eq:Dk-general}
\end{equation}
The sum runs over all computational-basis states with exactly $k$ spins in the state $\ket{\uparrow}$.

These states are orthonormal:
\begin{align}
    \bra{D_k}\ket{D_{k'}} = \delta_{k,k'}.
\end{align}

Moreover, each $\ket{D_k}$ is invariant under all permutations $P_\sigma$, because the sum in \eqref{eq:Dk-general} is simply permuted (reordered) by $P_\sigma$, leaving the set of terms unchanged.

Thus, the seven Dicke states
\begin{align}
\ket{D_0}, \ket{D_1}, \ket{D_2} \ket{D_3}, \ket{D_4}, \ket{D_5}, \ket{D_6}
\end{align}
span the fully symmetric subspace $\mathcal{H}_\text{sym}$ associated with the Young diagram $(6)$.

\onecolumngrid
\subsection{Explicit expansion of the Dicke states for $N=6$}

We now write each $|D_k\rangle$ explicitly in the computational basis.

\begin{align}
    \ket{D_0} = |\downarrow\downarrow\downarrow\downarrow\downarrow\downarrow\rangle.
\end{align}

\begin{align}
    \ket{D_1} = \frac{1}{\sqrt{6}}\Big(\ket{\uparrow\downarrow\downarrow\downarrow\downarrow\downarrow}+\ket{\downarrow\uparrow\downarrow\downarrow\downarrow\downarrow}
    +\ket{\downarrow\downarrow\uparrow\downarrow\downarrow\downarrow}+\ket{\downarrow\downarrow\downarrow\uparrow\downarrow\downarrow}
    +\ket{\downarrow\downarrow\downarrow\downarrow\uparrow\downarrow}
    +\ket{\downarrow\downarrow\downarrow\downarrow\downarrow\uparrow}
    \Big).
\end{align}

\begin{align}
    \ket{D_2} = \frac{1}{\sqrt{15}}\Big(&\ket{\uparrow\uparrow\downarrow\downarrow\downarrow\downarrow}
    +\ket{\uparrow\downarrow\uparrow\downarrow\downarrow\downarrow}
    +\ket{\uparrow\downarrow\downarrow\uparrow\downarrow\downarrow}
    +\ket{\uparrow\downarrow\downarrow\downarrow\uparrow\downarrow}
    +\ket{\uparrow\downarrow\downarrow\downarrow\downarrow\uparrow}
    +\ket{\downarrow\uparrow\uparrow\downarrow\downarrow\downarrow}
    +\ket{\downarrow\uparrow\downarrow\uparrow\downarrow\downarrow}
    +\ket{\downarrow\uparrow\downarrow\downarrow\uparrow\downarrow}
    +\ket{\downarrow\uparrow\downarrow\downarrow\downarrow\uparrow}\\\nonumber
    +&\ket{\downarrow\downarrow\uparrow\uparrow\downarrow\downarrow}
    +\ket{\downarrow\downarrow\uparrow\downarrow\uparrow\downarrow}
    +\ket{\downarrow\downarrow\uparrow\downarrow\downarrow\uparrow}
    +\ket{\downarrow\downarrow\downarrow\uparrow\uparrow\downarrow}
    +\ket{\downarrow\downarrow\downarrow\uparrow\downarrow\uparrow}
    +\ket{\downarrow\downarrow\downarrow\downarrow\uparrow\uparrow}
    \Big).
\end{align}

\begin{align}
    \ket{D_3} = \frac{1}{\sqrt{20}}\Big(&
    \ket{\uparrow\uparrow\uparrow\downarrow\downarrow\downarrow}
    +\ket{\uparrow\uparrow\downarrow\uparrow\downarrow\downarrow}
    +\ket{\uparrow\uparrow\downarrow\downarrow\uparrow\downarrow}
    +\ket{\uparrow\uparrow\downarrow\downarrow\downarrow\uparrow}
    +\ket{\uparrow\downarrow\uparrow\uparrow\downarrow\downarrow}
    +\ket{\uparrow\downarrow\uparrow\downarrow\uparrow\downarrow}
    +\ket{\uparrow\downarrow\uparrow\downarrow\downarrow\uparrow}
    +\ket{\uparrow\downarrow\downarrow\uparrow\uparrow\downarrow}
    +\ket{\uparrow\downarrow\downarrow\uparrow\downarrow\uparrow}\\\nonumber
    +&\ket{\uparrow\downarrow\downarrow\downarrow\uparrow\uparrow}
    +\ket{\downarrow\uparrow\uparrow\uparrow\downarrow\downarrow}
    +\ket{\downarrow\uparrow\uparrow\downarrow\uparrow\downarrow}
    +\ket{\downarrow\uparrow\uparrow\downarrow\downarrow\uparrow}
    +\ket{\downarrow\uparrow\downarrow\uparrow\uparrow\downarrow}
    +\ket{\downarrow\uparrow\downarrow\uparrow\downarrow\uparrow}
    +\ket{\downarrow\uparrow\downarrow\downarrow\uparrow\uparrow}
    +\ket{\downarrow\downarrow\uparrow\uparrow\uparrow\downarrow}
    +\ket{\downarrow\downarrow\uparrow\uparrow\downarrow\uparrow}\\\nonumber
    +&\ket{\downarrow\downarrow\uparrow\downarrow\uparrow\uparrow}
    +\ket{\downarrow\downarrow\downarrow\uparrow\uparrow\uparrow}
    \Big).
\end{align}

\begin{align}
    \ket{D_4} = \frac{1}{\sqrt{15}}\big(&
    \ket{\uparrow\uparrow\uparrow\uparrow\downarrow\downarrow}
    +\ket{\uparrow\uparrow\uparrow\downarrow\uparrow\downarrow}
    +\ket{\uparrow\uparrow\uparrow\downarrow\downarrow\uparrow}
    +\ket{\uparrow\uparrow\downarrow\uparrow\uparrow\downarrow}
    +\ket{\uparrow\uparrow\downarrow\uparrow\downarrow\uparrow}
    +\ket{\uparrow\uparrow\downarrow\downarrow\uparrow\uparrow}
    +\ket{\uparrow\downarrow\uparrow\uparrow\uparrow\downarrow}
    +\ket{\uparrow\downarrow\uparrow\uparrow\downarrow\uparrow}
    +\ket{\uparrow\downarrow\uparrow\downarrow\uparrow\uparrow}\\\nonumber
    +&\ket{\uparrow\downarrow\downarrow\uparrow\uparrow\uparrow}
    +\ket{\downarrow\uparrow\uparrow\uparrow\uparrow\downarrow}
    +\ket{\downarrow\uparrow\uparrow\uparrow\downarrow\uparrow}
    +\ket{\downarrow\uparrow\uparrow\downarrow\uparrow\uparrow}
    +\ket{\downarrow\uparrow\downarrow\uparrow\uparrow\uparrow}
    +\ket{\downarrow\downarrow\uparrow\uparrow\uparrow\uparrow}
    \big).
\end{align}

\begin{align}
    \ket{D_5} = \frac{1}{\sqrt{6}}\big(&
    \ket{\downarrow\uparrow\uparrow\uparrow\uparrow\uparrow}
    +\ket{\uparrow\downarrow\uparrow\uparrow\uparrow\uparrow}
    +\ket{\uparrow\uparrow\downarrow\uparrow\uparrow\uparrow}
    +\ket{\uparrow\uparrow\uparrow\downarrow\uparrow\uparrow}
    +\ket{\uparrow\uparrow\uparrow\uparrow\downarrow\uparrow}
    +\ket{\uparrow\uparrow\uparrow\uparrow\uparrow\downarrow}
    \big).
\end{align}

\begin{align}
    \ket{D_6} = \ket{\uparrow\uparrow\uparrow\uparrow\uparrow\uparrow}.
\end{align}

\twocolumngrid
\subsection{Permutation invariance of Dicke states}

For any permutation $\sigma \in S_6$, consider its action on $\ket{D_k}$:
\begin{align}
    P_\sigma \ket{D_k}
    = \frac{1}{\sqrt{\binom{6}{k}}}
    \sum_{\#\{\uparrow\}=k} P_\sigma \ket{s_1\cdots s_6}.
\end{align}
But as we permute the indices, we merely relabel the same set of basis states with $k$ spins up. Since the sum is over all such states with equal coefficients, $P_\sigma$ simply permutes the terms in the sum and thus leaves the state invariant:
\begin{align}
    P_\sigma \ket{D_k} = \ket{D_k},\  \forall \sigma\in S_6,\ \forall k.
\end{align}

Therefore, all Dicke states lie in the fully symmetric subspace
\begin{align}
    \mathcal{H}_\text{sym} = \{\ket{\psi}\in\mathcal{H}\,:\, P_\sigma\ket{\psi} = \ket{\psi},\ \forall\sigma\in S_6\},
\end{align}
and $\mathcal{H}_\text{sym}$ is exactly the $J=3$, Young-diagram $(6)$ sector.

\subsection{Other Young-diagram sectors and their dynamical closure}
We then discuss the other sectors that are not fully symmetric.

\subsection{Structure of the $(5,1)$, $(4,2)$, and $(3,3)$ sectors}

The sectors corresponding to Young diagrams $(5,1)$, $(4,2)$, and $(3,3)$ are not fully symmetric under all permutations. Instead, they transform according to nontrivial irreps of $S_6$.

For example:

\begin{itemize}
  \item The $(5,1)$ sector (associated with $J=2$) has multiplicity $f^{(5,1)} = 5$ and $SU(2)$ dimension $5$, giving a total of $25$ states. These states may be thought of as having ``one particle distinguished and five symmetric'' at the level of permutation symmetry.
  \item The $(4,2)$ sector (associated with $J=1$) has multiplicity $f^{(4,2)} = 9$ and $SU(2)$ dimension $3$, giving $27$ states.
  \item The $(3,3)$ sector (associated with $J=0$) has multiplicity $f^{(3,3)} = 5$ and $SU(2)$ dimension $1$, giving $5$ states.
\end{itemize}

These sectors are irreducible under the action of $S_6$, and each is invariant under all $P_\sigma$: if $\ket{\psi}$ lies in the sector corresponding to $\lambda$, then $P_\sigma \ket{\psi}$ is a linear combination of states in the same sector. However, in general
\begin{align}
    P_\sigma\ket{\psi} \neq \ket{\psi},
\end{align}
so these states are not symmetric in the sense of the fully symmetric sector.

\subsection{Example of a $(5,1)$-type state}
As a concrete example, consider the two-spin antisymmetric combination in spins $1$ and $2$:
\begin{align}
    \ket{\psi^-_{12}} = \frac{1}{\sqrt{2}}\left(\ket{\uparrow\downarrow} - \ket{\downarrow\uparrow}\right).
\end{align}
If we tensor this with some fully symmetric state of spins $3$--$6$, we obtain a state in the $(5,1)$ sector. For example,
\begin{align}
    \ket{\Psi_{(5,1)}} = \ket{\psi^-_{12}} \otimes \ket{\downarrow\downarrow\downarrow\downarrow}_{3-6}.
\end{align}
Under the transposition $(1\ 2)$, we have
\begin{align}
    P_{(1\,2)}\ket{\Psi_{(5,1)}} = -\ket{\Psi_{(5,1)}},
\end{align}
showing explicitly that such a state is not symmetric under all permutations. Nonetheless, the span of all $(5,1)$-type states is invariant under $S_6$.

\subsection{Two theorems}
Based on the preceding discussion, we can derive the following two theorems:

\begin{theorem}[Block-diagonal structure of $H$]
\label{th1}
Let $H$ satisfy $[H,P_\sigma]=0$ for all $\sigma\in S_6$. Then in the Schur–Weyl decomposition
\begin{align}
\mathcal{H} \cong V_3 \otimes W_{(6)}
\ \oplus\
V_2 \otimes W_{(5,1)}
\ \oplus\
V_1 \otimes W_{(4,2)}
\ \oplus\
V_0 \otimes W_{(3,3)},
\end{align}
$H$ is block-diagonal and can be written as
\begin{align}
H
=
H_{3}\otimes I_{(6)}
\ \oplus\
H_{2}\otimes I_{(5,1)}
\ \oplus\
H_{1}\otimes I_{(4,2)}
\ \oplus\
H_{0}\otimes I_{(3,3)},
\end{align}
where $H_J$ acts only on $V_J$ and $I_{\lambda}$ is the identity on $W_{\lambda(J)}$.
\end{theorem}

[Sketch of proof] This theorem can be proved by utilizing the Schur's lemma.

\begin{theorem}[Dynamical closure of each Young sector]
Let $H$ satisfy $[H,P_\sigma]=0$ for all $\sigma\in S_6$. Then each subspace
\begin{align}
\mathcal{H}_{(6)},\ 
\mathcal{H}_{(5,1)}, 
\mathcal{H}_{(4,2)},\ 
\mathcal{H}_{(3,3)},
\end{align}
corresponding respectively to the Young diagrams $(6)$, $(5,1)$, $(4,2)$, $(3,3)$, is invariant under the time evolution generated by $H$:
\begin{align}
e^{-iHt} \mathcal{H}_{\lambda} \subseteq \mathcal{H}_{\lambda},
\  \forall t\in\mathbb{R},\ \forall \lambda\in\{(6),(5,1),(4,2),(3,3)\}.
\end{align}
In particular, each sector is dynamically closed.
\end{theorem}
[Sketch of proof] This theorem can be proved by utilizing the block-diagonal nature of the Hamiltonian given in Theorem~\ref{th1}.

These two theorems can be directly generalized to general $N$ particles.

\section{Derivation for the infinite-$N$ classical dynamical system}\label{secA3}
\subsection{The classical  Eq.~\eqref{classicalEq}}
In the following, we outline the derivation of the classical Eq.\ref{classicalEq}. The evolution equations for $m^x(t)$, $m^y(t)$, and $m^z(t)$ can be obtained in a straightforward manner by substituting the collective spin operators into the original Lindblad equation (Eq.\ref{LindbladEq}) and then taking the trace. We take $m^x(t)$ as an example. we have
\begin{widetext}
    \begin{align}
        \frac{d}{dt} m^x &= \mathbf{Tr}[\frac{d\hat{\rho}}{dt}\hat{S}^x]
        = \underbrace{\mathbf{Tr}\big(-i[H,\hat{\rho}]\hat{S}^x\big)}_{\text{unitary term}} + \underbrace{\mathbf{Tr}\big(\frac{\kappa}{N}[2\hat{\sigma}^{-}\hat{\rho}\hat{\sigma}^{+}-\{\hat{\rho}, \hat{\sigma}^{+}\hat{\sigma}^{-}\}]\hat{S}^x \big)}_{\text{dissipation term}},
    \end{align}
where
\begin{align}
    &\text{unitary term} = \mathbf{Tr}\big(i\hat{\rho}H\hat{S}^x - iH\hat{\rho}\hat{S}^x\big)= \mathbf{Tr}\big(i\hat{\rho}H\hat{S}^x - i\hat{\rho}\hat{S}^x H\big)= \mathbf{Tr}\big(i\hat{\rho}[H,\hat{S}^x]\big)\nonumber\\
    &= \mathbf{Tr}\Bigg(\hat{\rho} \bigg[ i\frac{N\Gamma}{2}[1-\cos{(\omega t)}]\big(\frac{-2i\hat{S}^z}{N}\big) + iNJ_y \hat{S}^y\big(\frac{-2i\hat{S}^z}{N}\big) + iNJ_y \big(\frac{-2i\hat{S}^z}{N}\big)\hat{S}^y + iNJ_z \hat{S}^z\big(\frac{2i\hat{S}^y}{N}\big) + iNJ_z \big(\frac{2i\hat{S}^y}{N}\big)\hat{S}^z \bigg]\Bigg)\nonumber\\
    &= \mathbf{Tr}\Bigg(\hat{\rho} \bigg[ \Gamma[1-\cos{(\omega t)}]\hat{S}^z + 2J_y\hat{S}^y\hat{S}^z+ 2J_y\hat{S}^z\hat{S}^y - 2J_z\hat{S}^z\hat{S}^y- 2J_z\hat{S}^y\hat{S}^z\bigg]\Bigg)\nonumber\\
    &=\Gamma[1-\cos{(\omega t)}]m^z + 4(J_y-J_z)m^ym^z,\\
    &\text{dissipation term} = \mathbf{Tr}\Bigg(\frac{2\kappa}{N}\hat{\sigma}^{-}\hat{\rho}\hat{\sigma}^{+}\hat{S}^x-\frac{\kappa}{N}\hat{\rho}\hat{\sigma}^{+}\hat{\sigma}^{-}\hat{S}^x-\frac{\kappa}{N}\hat{\sigma}^{+}\hat{\sigma}^{-}\hat{\rho}\hat{S}^x \Bigg)\nonumber\\
    &= \frac{\kappa}{N}\mathbf{Tr}\Bigg(2\hat{\rho}\hat{\sigma}^{+}\hat{S}^x\hat{\sigma}^{-}-\hat{\rho}\hat{\sigma}^{+}\hat{\sigma}^{-}\hat{S}^x-\hat{\rho}\hat{S}^x\hat{\sigma}^{+}\hat{\sigma}^{-} \Bigg)\nonumber\\
    &= \frac{\kappa N}{4}\mathbf{Tr}\Bigg(2\hat{\rho}(\hat{S}^x +i\hat{S}^y)\hat{S}^x(\hat{S}^x -i\hat{S}^y)-\hat{\rho}(\hat{S}^x +i\hat{S}^y)(\hat{S}^x -i\hat{S}^y)\hat{S}^x-\hat{\rho}\hat{S}^x(\hat{S}^x +i\hat{S}^y)(\hat{S}^x -i\hat{S}^y) \Bigg)\nonumber\\
    &= \kappa m^x m^z.
\end{align}
The last equality is obtained by inserting the commutation relations:
\begin{align}
    [\hat{S}^x, \hat{S}^x-i\hat{S}^y] &= -i[\hat{S}^x, \hat{S}^y] = 2\hat{S}^z/N,\nonumber\\
    [\hat{S}^x,(\hat{S}^x +i\hat{S}^y)(\hat{S}^x -i\hat{S}^y)] &= \frac{2}{N}(\hat{S}^x\hat{S}^z-\hat{S}^z\hat{S}^x +i\hat{S}^y\hat{S}^z+i\hat{S}^z\hat{S}^y).
\end{align}
Similar calculations can be performed for $m^y(t)$ and $m^z(t)$. Then we obtain the classical Eq.~\eqref{classicalEq} in the main text.

\end{widetext}

\subsection{Comparison between different numerical methods}
\begin{figure}[t]
\centering
\includegraphics[width=0.48\textwidth]{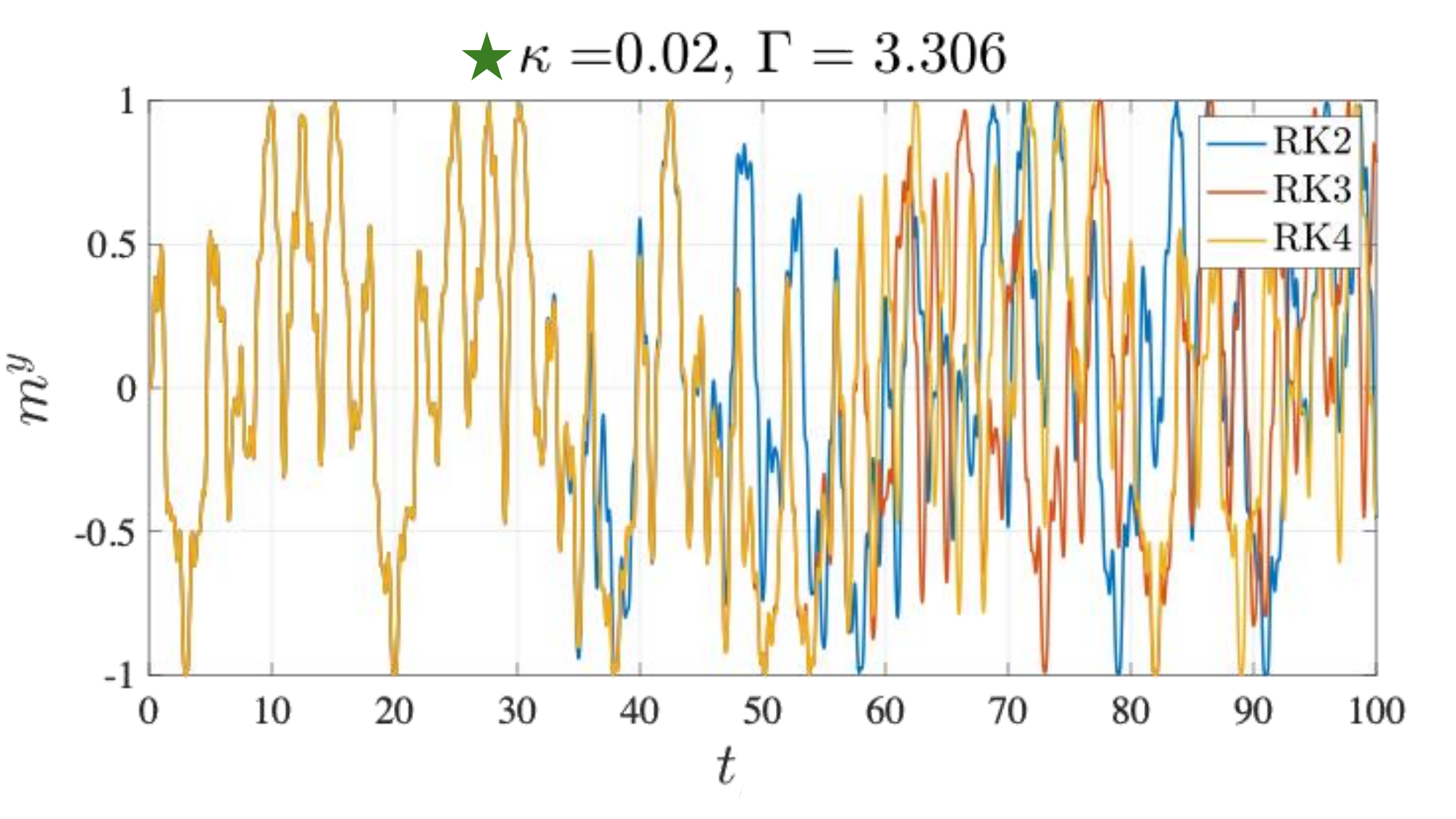}
\caption{Dynamics in the first $100$ periods obtained in the chaotic phase using different numerical integration schemes. The three curves show the $m^y(t)$ component of the solutions to Eq.~\eqref{classicalEq} evolved using the RK2, RK3, and RK4 methods, labeled by blue, red and yellow, respectively.}
\label{RK234}
\end{figure}
To further illustrate that the system’s evolution orbits are ill-defined in the chaotic phase, we perform time evolution of Eq.\ref{classicalEq} using different numerical integration methods, as shown in Fig.\ref{RK234}. In particular, we compare between RK2, RK3, and RK4 methods, and focus on $m^y(t)$. At early times, errors induced by the MLE have not yet accumulated into observable deviations, so the different methods yield similar dynamics. However, at later times, as the positive MLE takes effect, the observable quantities begin to diverge quantitatively. This further demonstrates the highly unpredictable nature of time evolution in the chaotic phase. It is also worth noting that from RK2 to RK3 and then to RK4, the numerical accuracy improves, meaning that the errors or perturbations introduced by the numerical integration become progressively smaller. As a result, for a fixed MLE it takes increasingly longer for these numerical errors to grow into macroscopically observable deviations. As shown in the figure, the orbit computed using the RK2 method begins to deviate from those of RK3 and RK4 around $t\approx35$, while the deviation between RK3 and RK4 becomes visible around $t\approx55$.

\section{The dynamics of the classical counterpart of Fig.~\ref{finiteN_onSphere}}
In Sec.~\ref{sec_finiteNandinfiniteN} we discussed the quantum evolution of the system in the classical chaotic phase for six different choices of initial states in Fig.~\ref{finiteN_onSphere}. For comparison, we show the corresponding classical orbit over the sphere with the same parameters and initial conditions here, together with the $x$-component of the orbit and the Fourier transform in early($0-50$) and late($150-200$) time. The orbits for the six initial states exhibit obvious chaos as expected by the positive MLEs.

\begin{figure*}[t]
\centering
\includegraphics[width=0.85\textwidth]{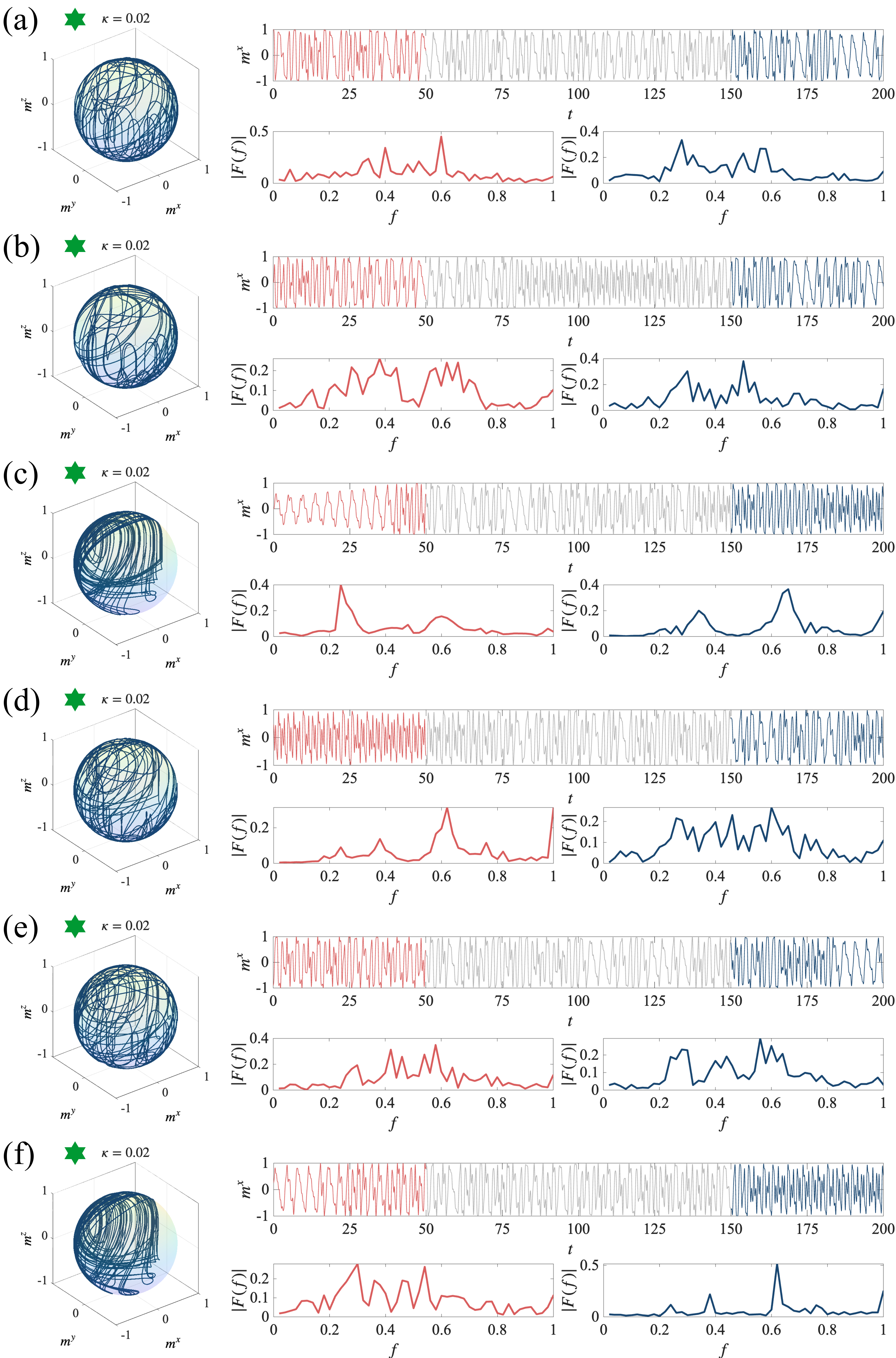}
\caption{\label{sphereorbit_M_FFt_2}The classical evolution at $\kappa = 0.02$ and $\Gamma = 3.306$. Six different initial states are chosen to be the same as in Fig.~\ref{finiteN_onSphere}.
The left panels show the trajectories of $\mathbf{m}(t)$ for $t \in [150, 200]$.
The right panels show the $x$-component of the orbit and their Fourier transform in early and late time, respectively.}
\end{figure*}

\section{The dynamics of the quantum counterpart of Fig.~\ref{sphereorbit_M_FFt}}

\begin{figure*}[t]
\centering
\includegraphics[width=0.85\textwidth]{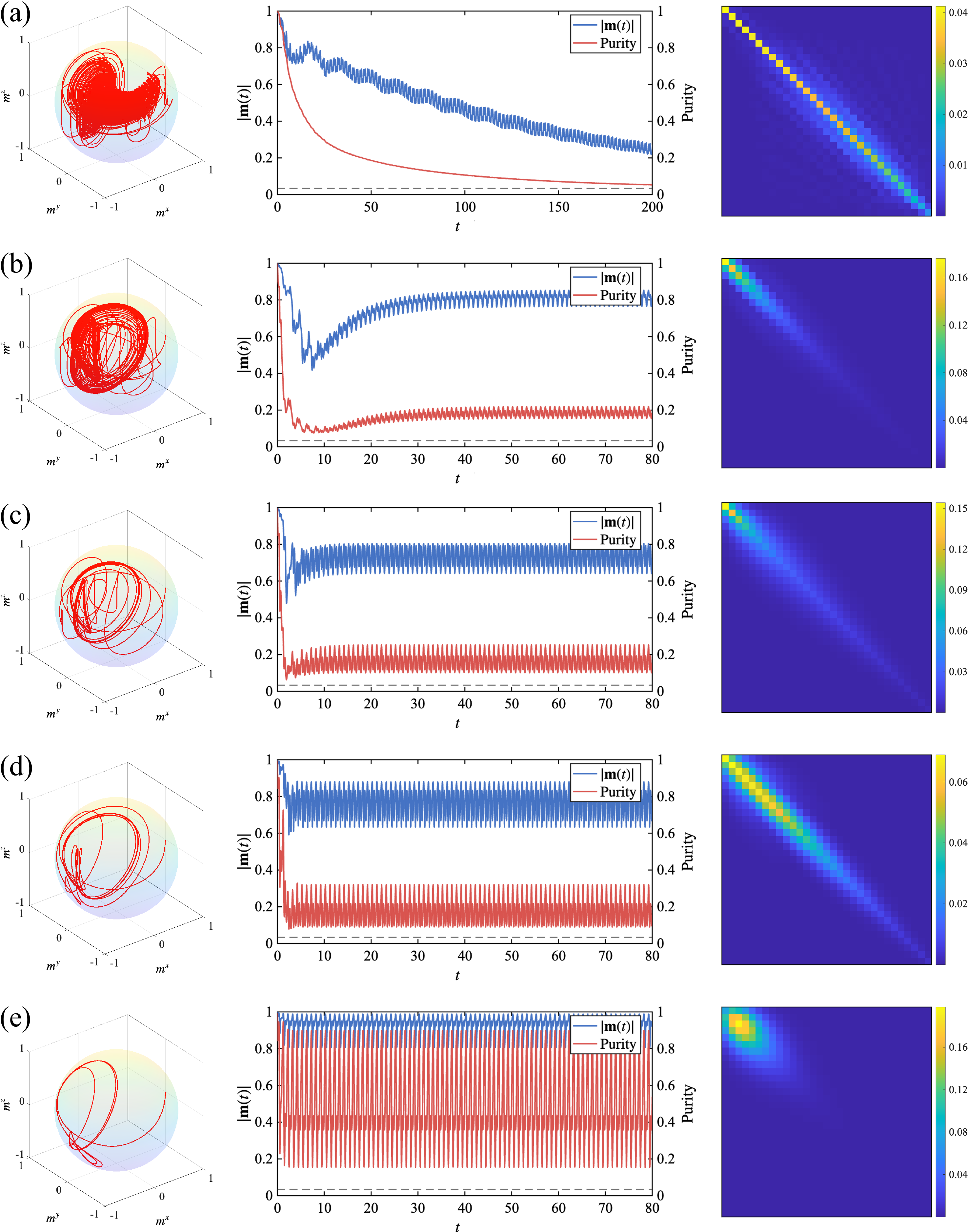}
\caption{\label{finiteN_onSphere_2}The quantum evolution at $\Gamma = 8.427$. Initial states are chosen to be $x$-polarized, the same as in Fig.~\ref{sphereorbit_M_FFt}.
The left panels show the trajectories of $\mathbf{m}(t)$ for $t \in [0, 80]$.
The middle panels display the time evolution of $|\mathbf{m}(t)|$ and the purity $\mathbf{Tr}[\hat{\rho}^2(t)]$, where the black dashed line denotes the purity of a maximally mixed state.
The rightmost panels show snapshots of the density matrix at $t = 80$ for system size $L = 30$.}
\end{figure*}
In this section, we present the quantum dynamics corresponding to the same parameter sets as in Fig.~\ref{sphereorbit_M_FFt}.

Figure~\ref{sphereorbit_M_FFt}(a) shows the chaotic case. The trajectory exhibits clear irregularity, and the norm of the polarization vector $|\mathbf{m}(t)|$ gradually decreases toward zero over time. The purity $\mathbf{Tr}[\hat{\rho}^2(t)]$ rapidly approaches the lower bound set by a maximally mixed state. The density matrix displays nonzero elements mainly concentrated along the diagonal, indicating that the system diffusively explores the entire Hilbert space—signifying the onset of quantum chaos.

Figure~\ref{sphereorbit_M_FFt}(b) corresponds to a quasiperiodic case. Although the system has a positive MLE, the interbranch periodicity dominates over the intrabranch chaotic fluctuations. Moreover, because of the relatively small system size ($L=30$), finite-size effects become significant: both $|\mathbf{m}(t)|$ and $\mathbf{Tr}[\hat{\rho}^2(t)]$ remain at finite values instead of decaying to zero. The density matrix shows partial localization compared to the chaotic case, but the localization is not strong enough to ensure convergence between the quantum and classical orbits.

Figure~\ref{sphereorbit_M_FFt}(c) corresponds to a period-2 orbit, where the MLE is negative, indicating a stable phase. Both $|\mathbf{m}(t)|$ and $\mathbf{Tr}[\hat{\rho}^2(t)]$ remain finite, and the density matrix exhibits a pronounced delocalized structure. In this regime, the discrepancy between the quantum and classical trajectories originates from the suppression of nontrivial periodic responses in the finite-size system.

Figures~\ref{sphereorbit_M_FFt}(d) and \ref{sphereorbit_M_FFt}(e) both correspond to the period-1 phase. The difference lies in the number of classical attractors: two in Fig.~\ref{sphereorbit_M_FFt}(d) and one in Fig.~\ref{sphereorbit_M_FFt}(e) (see the discussion in connection with Fig.~\ref{010bifurcationLinecut}). In both cases, $|\mathbf{m}(t)|$ stabilizes around a finite value close to unity, but $\mathbf{Tr}[\hat{\rho}^2(t)]$ behaves very differently. With two attractors, $\mathbf{Tr}[\hat{\rho}^2(t)]$ quickly saturates near 0.2, while with a single attractor, it exhibits large oscillations between 0.2 and 0.9. The density matrices also show distinct features: in the two-attractor case, although the main contributions of the nonzero elements are localized, many small but finite elements still spread across the Hilbert space, indicating a partially delocalized nature. It appears as if the system “senses’’ the presence of the other attractor during its evolution. In contrast, when there is only one global attractor, the density matrix becomes strongly localized, and the quantum and classical orbits yield consistent dynamics.

\section{Fractal of the regions of attractors}\label{secA4}

\begin{figure*}[t]
\centering
\includegraphics[width=1\textwidth]{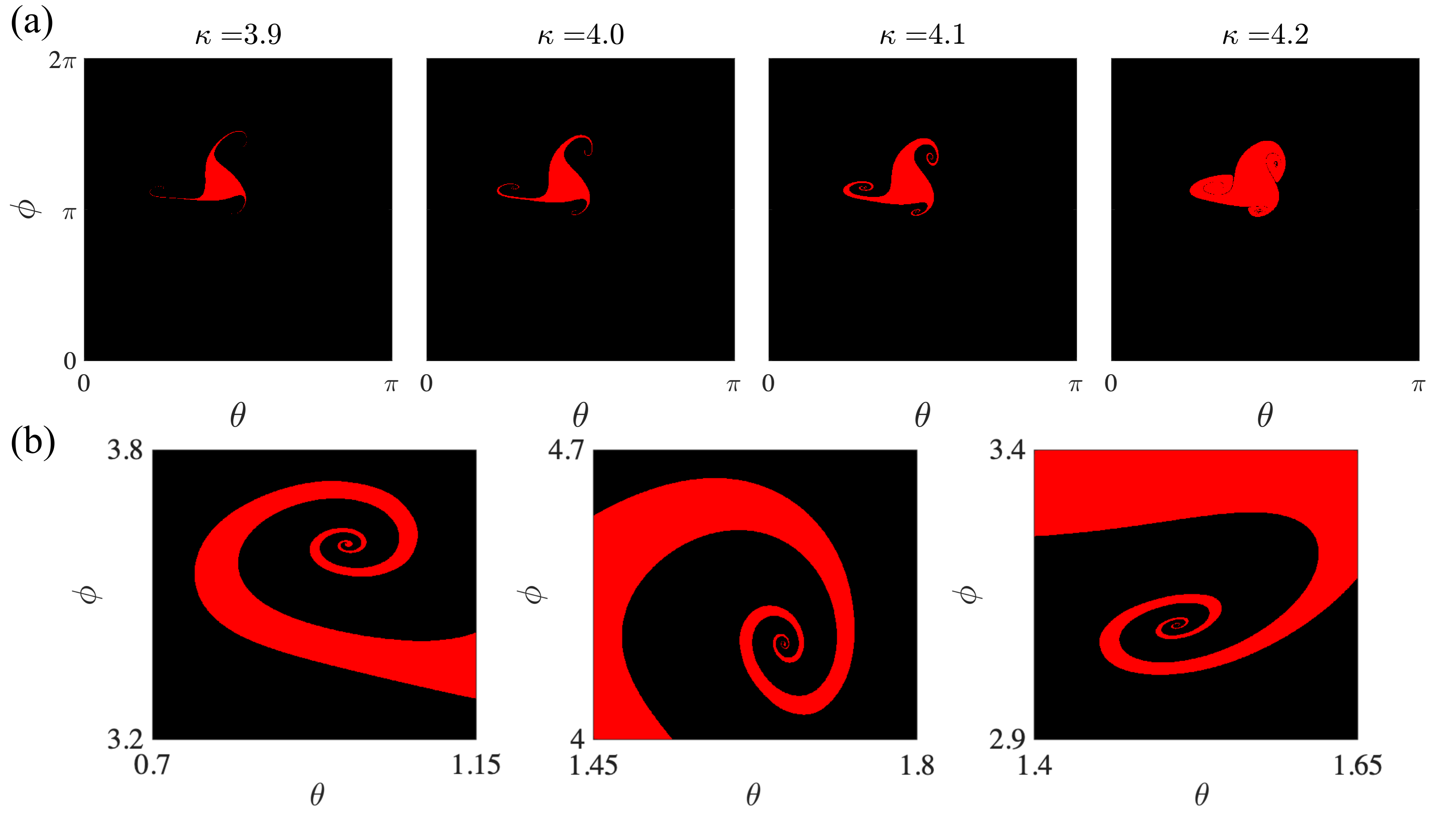}
\caption{\label{regionofattraction2}(a) The evolution of the regions of attractors over the sphere as a function of $\kappa$. The value of $\kappa$ ranges from $3.9$ to $4.2$. We use the same black and red coloring scheme as in the corresponding parameter range of Fig.~\ref{010bifurcationLinecut}(a). (b) Zoom-in plots for the three fractal rotational regions for $\kappa=4.1$.}
\end{figure*}

\begin{figure*}[t]
\centering
\includegraphics[width=1.03\textwidth]{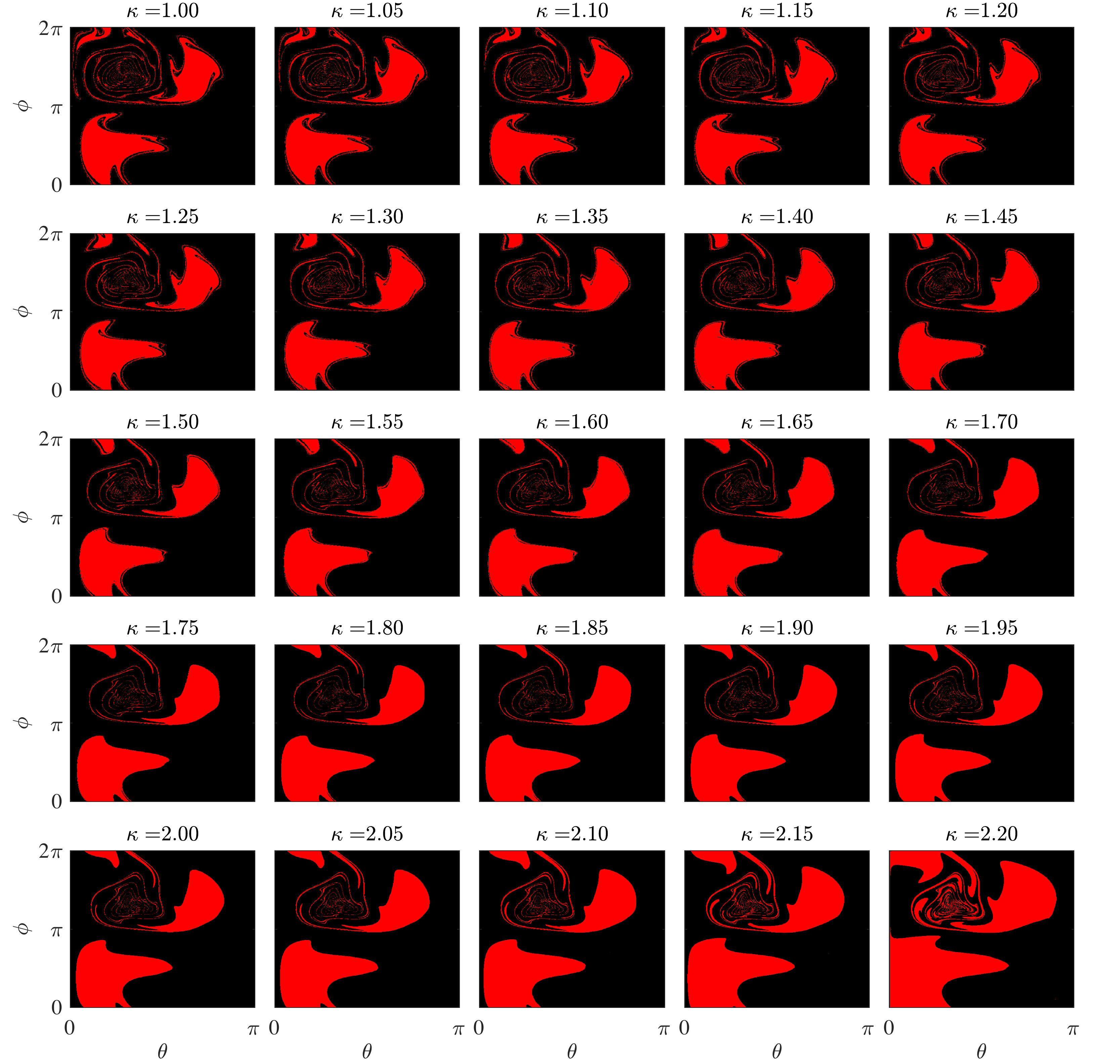}
\caption{\label{010quasi_regionofattraction}The evolution of the regions of attractors over the sphere as a function of $\kappa$. The value of $\kappa$ ranges from $1.00$ to $2.20$. We use the same black and red coloring scheme as in the corresponding parameter range of Fig.~\ref{010bifurcationLinecut}(a) to distinguish the regions of attractors of the quasiperiodic branch and others.}
\end{figure*}

In Sec.\ref{sec_bifurctaion}, we examined the evolution of the regions of attractors for $\kappa > 5.0$, where a continuous deformation of the regions leads to the apparent discontinuities observed in the bifurcation diagram, Fig.\ref{010bifurcationLinecut}(a). However, a much more intriguing deformation process emerges at lower values of $\kappa$, particularly around $\kappa \approx 4.0$, with the same driving strength $\Gamma=8.427$. Although the system in this regime also hosts two stable attractors—resulting in two corresponding regions of attractors on the sphere—their shapes become highly irregular and display strikingly singular structures known as fractals. 

In Fig.~\ref{regionofattraction2}(a), we present the evolution of the regions of attractors as $\kappa$ varies from $3.9$ to $4.2$. A dramatic twisting of the red region’s three ``corners" can be clearly observed, ultimately giving rise to fractal boundaries. One of the defining features of fractals is self-similarity. By zooming into the three twisted regions in the $\kappa = 4.1$ case in Fig.~\ref{regionofattraction2}(b), we observe evident self-similar patterns and infinitely rotations, indicating that the region boundaries indeed exhibit fractal characteristics at this parameter regime.

Furthermore, in Fig.~\ref{010quasi_regionofattraction}, we analyze the region of attraction associated with the quasiperiodic branch, shown in black, while the others are colored red. Here, even richer deformation behaviors are revealed. The appearance of self-similar fractal structures, along with infinitely nested rotations, suggests the possible presence of a Smale horseshoe structure ~\cite{de1999pruning,igra2025perioddoublingcascadesinvariantsbraided}, hinting at a deeper and more complex underlying dynamical mechanism.

\section{Period-doubling Bifurcation}{\label{001bifurcationsection}}
\begin{figure*}[ht]
\centering
\includegraphics[width=1\textwidth]{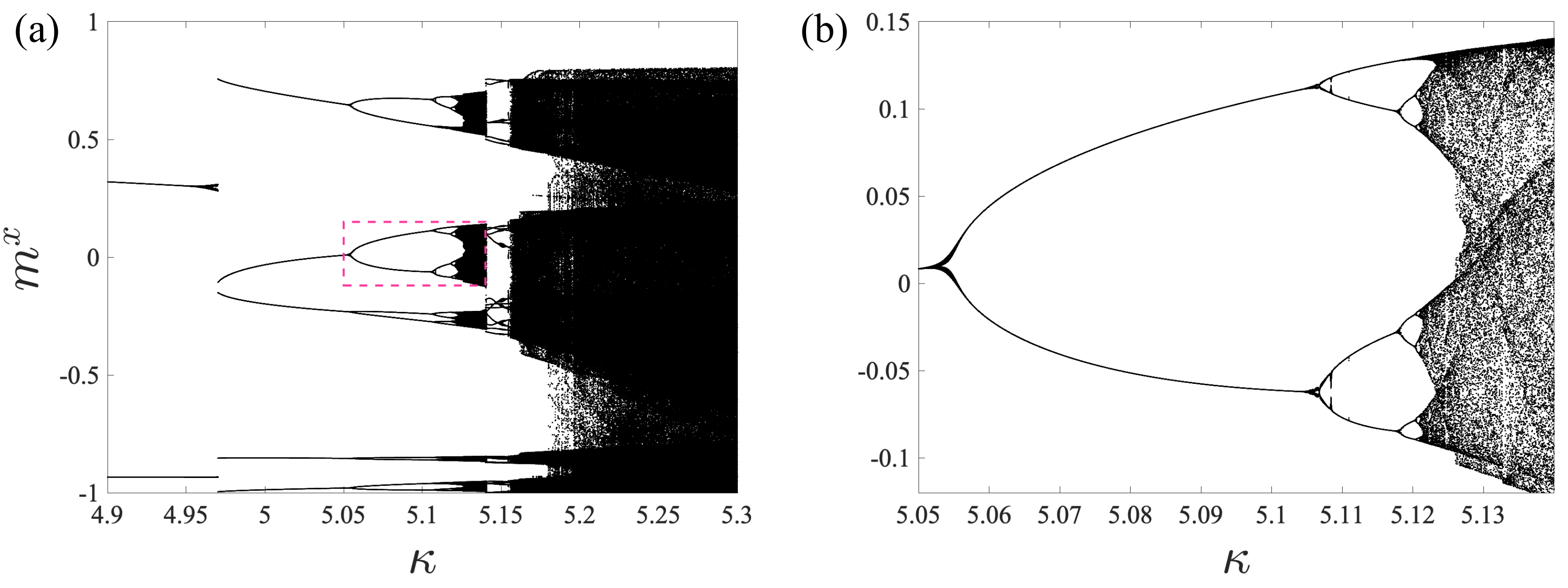}
\caption{\label{001perioddoublingbifurcation}Period-doubling bifurcation of $m^x(t)$ for interaction strength $(J_x, J_y, J_z) = (0, 0, 1)$. Panel (a) illustrates a typical route to chaos via a sequence of period-doubling bifurcations. As $\kappa$ increases, the system undergoes successive bifurcations starting from a simple periodic orbit, each doubling the period of the previous one, eventually leading to the onset of chaotic dynamics. The region enclosed by the red dashed box highlights a representative period-doubling bifurcation. In panel (b), we zoom into this region, where the successive period-doubling process becomes clearly visible. The calculations are performed at a fixed driving strength of $\Gamma = 18.409$, and the initial state is set to be $x$-polarized.}
\end{figure*}
We now turn to a more fundamental and intriguing feature observed in our system: the period-doubling bifurcation~\cite{sander2010connectingperioddoublingcascadeschaos}. The interaction strength is set to be $(J_x, J_y, J_z) = (0, 0, 1)$ here.

A period-doubling bifurcation occurs when a small change in a system parameter destabilizes a periodic orbit and gives rise to a new one with twice the period of the original. This process can repeat itself, resulting in a cascade of bifurcations where the period continues to double successively. Such period-doubling cascades are a hallmark route to chaos and signal the emergence of highly complex, sensitive dependence on initial conditions.

In Fig.~\ref{001perioddoublingbifurcation}(a), we present a representative example of the period-doubling bifurcation. The driving strength is fixed at $\Gamma = 18.409$. Starting from $\kappa \approx 5.05$, the system undergoes successive period-doubling transitions, beginning from a stable orbit with period $5$ and progressing toward chaos. Notably, around $\kappa \approx 5.15$, a narrow stable window is embedded in chaotic phases—a familiar feature we have mentioned before in Fig.~\ref{010bifurcationLinecut}. The red dashed box in Fig.~\ref{001perioddoublingbifurcation}(a) marks a typical region where the period-doubling bifurcation shows up. In Fig.~\ref{001perioddoublingbifurcation}(b), we zoom into this region, where the successive doublings of the orbit period become clearly visible. The intervals between successive bifurcations shrink rapidly, suggesting an accelerated approach toward chaos as $\kappa$ increases. This is a key feature of the period-doubling route to chaos—one of the canonical pathways by which dynamical systems develop complex, chaotic behavior. Indeed, the progressively shortening intervals between successive period doublings can be quantitatively described and verified by the first Feigenbaum constant, or simply the Feigenbaum constant $\delta$~\cite{feigenbaum1978quantitative}. The constant $\delta$ is defined as the limiting ratio of each bifurcation interval to the next between consecutive period doublings. Its approximate value, $\delta \approx 4.669$, reflects the universal scaling behaviour observed in a wide class of nonlinear dynamical systems undergoing period-doubling cascades on the way to chaos. In the calculation shown in Fig.~\ref{001perioddoublingbifurcation}, we numerically estimate the ratio of successive bifurcation intervals and obtain a value of approximately $4.633$, which is in close agreement with the Feigenbaum constant. This supports the interpretation that the observed bifurcation cascade indeed follows the universal period-doubling bifurcation.

\bibliography{ref}

\end{document}